\documentclass[pdftex, 12pt, a4paper]{article}
\usepackage[pdftex]{color, graphicx}
\usepackage[margin=0.8in]{geometry}
\usepackage[english]{babel}
\usepackage[justification=justified,font=small,labelfont=small]{caption}
\usepackage{subcaption}
\usepackage{appendix, parskip, inputenc, tocbibind, url}
\usepackage{wrapfig, enumitem, multirow, tabularx}
\usepackage{amsmath, amssymb, amsthm, braket, empheq, fixltx2e, commath, mathtools}
\usepackage{cleveref}
\usepackage{ragged2e}
\usepackage{gensymb}
\usepackage{float}
\usepackage{cite}

\makeatletter
\newtheoremstyle{indented}
{6pt}
{}
{\addtolength{\@totalleftmargin}{3.5em}
   \addtolength{\linewidth}{-3.5em}
   \parshape 1 3.5em \linewidth}
{}
{\bfseries}
{.}
{.5em}
{}
\newtheoremstyle{indenteditalic}
{6pt}
{}
{\em
   \addtolength{\@totalleftmargin}{3.5em}
   \addtolength{\linewidth}{-3.5em}
   \parshape 1 3.5em \linewidth}
{}
{\bfseries}
{.}
{.5em}
{}
\newtheoremstyle{proof}
{6pt}
{}
{\addtolength{\@totalleftmargin}{3.5em}
   \addtolength{\linewidth}{-3.5em}
   \parshape 1 3.5em \linewidth}
{}
{\em}
{.}
{.5em}
{}
\makeatother

\theoremstyle{indenteditalic}

\numberwithin{thm}{section}

\theoremstyle{indenteditalic}

\theoremstyle{indenteditalic}

\theoremstyle{indented}

\theoremstyle{indented}

\theoremstyle{proof}

\theoremstyle{proof}

\theoremstyle{proof}

\let\bg\begin

\newcommand{\bal}{\bg{align}}

\newcommand{\bsubs}{\bg{subequations}}
\newcommand{\esubs}{\end{subequations}}
\newcommand{\barr}{\begin{array}}
\newcommand{\earr}{\end{array}}

\let\tn\textnormal

\let\tb\textbf
\let\tsups\textsuperscript
\let\tsubs\textsubscript

\let\mbb\mathbb
\let\mcal\mathcal

\let\lt\left
\let\rt\right
\let\fr\frac

\let\rrr\rightarrow

\let\h\hat

\let\tld\tilde

\let\p\partial

\let\kt\ket
\let\brk\braket

\let\a\alpha
\let\b\beta
\let\Gm\Gamma
\let\gm\gamma
\let\de\delta
\let\De\Delta
\let\O\Omega

\let\s\sigma
\let\th\theta
\let\eps\epsilon

\let\n\nu
\let\Ph\Phi
\let\ph\phi

\let\Ps\Psi
\let\ps\psi
\let\lm\lambda

\let\t\tau
\let\hb\hbar
\let\rh\rho

\let\k\kappa
\let\th\theta

\renewcommand{\d}{\tn{d}}

\newcommand{\half}{\fr{1}{2}}
\newcommand{\hH}{\h{H}}
\newcommand{\hHe}{\h{H}_\tn{e}}
\newcommand{\hHp}{\h{H}_\tn{p}}
\newcommand{\hHint}{\h{H}_\tn{int}}
\newcommand{\ha}{\h{A}}
\newcommand{\had}{\h{A}^{\dagger}}
\newcommand{\ktvac}{\kt{\tn{vac}}}

\newcommand{\ktPs}{\kt{\Ps}}

\newcommand{\ktPst}{\kt{\Ps(t)}}

\newcommand{\SE}{Schr\"{o}dinger equation}
\newcommand{\ang}{\r{A}}
\newcommand{\ahelix}{$\a$-helix}
\newcommand{\Romanone}{\uppercase\expandafter{\romannumeral1}}

\newcommand{\Romantwo}{\uppercase\expandafter{\romannumeral2}}

\newcommand{\Eb}{E_\tn{b}}
\newcommand{\td}{\t_\tn{d}}

\newcommand{\zeroN}{\sum_{n=0}^N}
\newcommand{\zeroNmone}{\sum_{n=0}^{N-1}}
\newcommand{\oneNmone}{\sum_{n=1}^{N-1}}

\newcommand{\dotpsn}{\dot{\psi}_n}

\newcommand{\dotu}{\dot{u}}
\newcommand{\ddotu}{\ddot{u}}

\newcommand{\chir}{\chi_r}
\newcommand{\chil}{\chi_l}

\newcommand{\ansq}{\abs{\a_n}^2}
\newcommand{\anponesq}{\abs{\a_{n+1}}^2}
\newcommand{\anmonesq}{\abs{\a_{n-1}}^2}
\newcommand{\psnsq}{\abs{\ps_n}^2}
\newcommand{\psnponesq}{\abs{\ps_{n+1}}^2}
\newcommand{\psnmonesq}{\abs{\ps_{n-1}}^2}

\newcommand{\sech}{\tn{sech}}
\newcommand{\arsech}{\tn{arsech}}
\newcommand{\arsinh}{\tn{arsinh}}

\newcommand{\xmax}{x_\tn{max}}

\newcommand{\sgn}{\tn{sgn}}
\newcommand{\maxpsnsq}{\max\abs{\ps_n}^2}
\newcommand{\bareps}{\bar{\eps}}
\newcommand{\epscomb}{\eps_{\tn{comb}}}
\newcommand{\Ac}{A_\tn{c}}
\newcommand{\Am}{A_\tn{m}}
\newcommand{\Vc}{V_\tn{c}}
\newcommand{\Vm}{V_\tn{m}}
\newcommand{\thc}{\th_\tn{c}}


\bg{document}
\pagestyle{plain}
\setlength{\baselineskip}{14.4pt}

\begin{titlepage}
\begin{center}

\vspace*{60pt}
\LARGE{\tb{A generalised Davydov-Scott model for polarons in linear peptide chains}}

\vspace{14.4pt}
\large{April 2017}

\vspace{60pt}
\large{\textsc{J. Luo}*}

\large{\textsc{B. M. A. G. Piette}$\dagger$}

\vfill
\large{*jingxi.luo@durham.ac.uk}

\large{$\dagger$b.m.a.g.piette@durham.ac.uk}

\vspace{14.4pt}
\large{Department of Mathematical Sciences, Durham University}

\end{center}
\end{titlepage}\begin{titlepage}
\begin{center}

\begin{minipage}{\textwidth}

\vspace*{120pt}
\begin{flushleft}
\Large{\textsc{Abstract}}
\end{flushleft}

\vspace{14.4pt} 
We present a one-parameter family of mathematical models describing the 
dynamics of polarons in periodic structures, such as linear polypeptides, 
which, by tuning the model parameter, can be reduced to
the Davydov or the Scott model. We describe the physical
significance of this parameter and, in the continuum limit, we derive
analytical solutions which represent stationary polarons. On a
discrete lattice, we compute stationary polaron solutions
numerically. We investigate polaron propagation induced by several
external forcing mechanisms. We show that an electric field consisting
of a constant and a periodic component can induce polaron motion with
minimal energy loss. We also show that thermal fluctuations can
facilitate the onset of polaron motion. Finally, we discuss the
bio-physical implications of our results.

\vspace{14.4pt}
PACS number(s): 71.38.-k, 63.20.kd, 05.60.-k, 05.40.-a

\end{minipage}
\end{center}
\end{titlepage}	\section{Introduction} \label{intro}

The polaron, a quasi-particle formed by the coupling of an electron to a vibrating lattice, was first theorised by L. D. Landau in 1933 \cite{Landau1933}. In essence, polaron formation is a process of electron self-trapping. The presence of the electron causes localised distortions in the natural vibrational mode of the lattice, a.k.a. the lattice phonon. In return, if the electromagnetic interaction between the electron and lattice is appropriate, then the phonon distortions can lower the potential well for the electron, thus trapping the electron. 

Some twenty years after the inception of the polaron concept, a mathematical description of it was formalised by H. Fr\"{o}hlich \cite{Frohlich1952} and subsequently T. Holstein \cite{Holstein1959a,Holstein1959b}. Since then, properties of the Fr\"ohlich-Holstein polaron have been well studied, with some authors hypothesising an application of dynamical polarons as electron transporters in conductive material \cite{Heeger1988,Voulgarakis2000,Brizhik2003}. In the 1970s, A. S. Davydov used the basis of polaron theory to explain some biological processes \cite{Davydov1979}. Specifically he proposed that, in an $\a$-helical protein, a certain intramolecular oscillator can interact with the peptide chain in a way similar to an electron interacting with a crystal lattice. Davydov suggested that this interaction could lead to the localisation and propagation of vibrational energy in the \ahelix. 
Later, A. C. Scott modified Davydov's theory, taking into account the internal geometry of peptide units \cite{Scott1992}.
Some authors argued that, given the polarisability of peptide units, electron self-trapping is also possible in proteins, and it can be described by the same mathematical model that Davydov and Scott used \cite{Chuev1993, Conwell2000}. It should therefore be possible that polaronic transport of electrons may take place in proteins, too. Recently, L. S. Brizhik \emph{et al.} reported on the properties of static and dynamical polarons in simple molecular chains, and adverted to the applicability of their results to electron transport in biomolecules such as proteins \cite{Brizhik2008,Brizhik2010,Brizhik2014}. Their studies were based on the Davydov-Scott model. 

In the current study, we propose a generalisation to the Davydov-Scott model, and use it to explore the properties of polarons in a linear peptide chain. In the generalised model, there is an extra parameter which represents the extent to which the electron-polypeptide interaction is spatially symmetric. In \cref{section2}, we describe our model and explain why the extra parameter is necessary. We also give physical interpretations of all other parameters in the model, justifying the choices of their values where possible. Then, we derive a set of coupled dynamical equations which govern the electron and phonon parts of the polaron, as well as how they interact. In \cref{section3} we look at solutions to our equations which are stationary, and thus deduce properties of static polarons admissible by our model, such as the polaron's binding energy. The process of solving the equations is carried out analytically as well as numerically. By the former approach, a closed-form expression for the solution is found, but its use is limited, because the solution process involves a few approximations and simplifying assumptions. By the numerical approach, no convenient expression for the solution is possible, but the method solves the equations directly without simplifications. We compare the results produced by the two different methods. 

\Cref{section4} concerns dynamical polarons. We discover that it is possible to use a suitable external forcing to displace the stationary polaron, and to sustain its motion in such a way that its energy remains highly stable. We investigate how the polaron's motion depends upon our forcing parameters. We use only numerical methods to obtain our results in \cref{section4}, as well as those in \cref{section5}, where we consider how the polaron's motion is affected by temperature of the environment. For this part, the external forcing from \cref{section4} remains in place, but we also utilise a parameter which controls the magnitude of the thermal effect. To account for the random nature of thermal fluctuations, we repeat each numerical simulation many times over, taking the average of the results. Finally, we conclude by discussing the physical realisabililty of our mathematical model, particularly how the external forcing which we study in \cref{section4} may be realised. We also briefly discuss the generalisability of our model to studying electron transport by polarons in $\a$-helices.


	\section{The model and dynamical equations} \label{section2}

In both Davydov's and Scott's models, the Hamiltonian for a system of excitons interacting with one-dimensional lattice phonons is written in Fr\"{o}hlich-Holstein form $\hH = \hHe + \hHp + \hHint$, where $\hHe, \hHp$ and $\hHint$ represent energy contributions from the exciton, phonon and interaction parts, respectively \cite{Frohlich1952,Holstein1959a,Davydov1982,Davydov1991,Scott1992}. We adopt this Hamiltonian for our model, and following \cite{Brizhik2008,Brizhik2010,Brizhik2014} we consider an additional \emph{external Hamiltonian}, $\hH_\tn{ext}$, so that our Hamiltonian takes the form
	\bal
	\hH = \hHe + \hHp + \hHint + \hH_\tn{ext},
	\end{align}
where $\hHe$ describes a tight-binding electron, the stretching and compressing of hydrogen bonds in the peptide chain are phonon oscillations described by $\hHp$, $\hHint$ accounts for the electron-phonon interaction, and $\hH_\tn{ext}$  represents the effect of an external electric field. We assume that the peptide chain consists of $N+1$ identical units and $N$ identical hydrogen bonds. In the tight-binding approximation, we have
	\bal
	\hHe = \sum_{n=0}^N J_0 \had_n \ha_n - \sum_{n=0}^{N-1} J_1 \lt(\had_{n+1}\ha_n + \had_n\ha_{n+1}\rt). \label{DSham_e}
	\end{align}
The subscript $n$ in \cref{DSham_e} labels peptide units, which are the unit cells of our lattice. $\had_n$ and $\ha_n$ are local electron creation and annihilation operators, respectively. $J_0$ is the potential energy of a localised electron. Modelling each unit as a point-dipole, we assume the nearest-neighbour dipole interaction energy is a constant and write it as $-J_1$ \cite{Eilbeck1984,Lehninger1993,Barford2007}. The external Hamiltonian,
	\bal
	\hH_\tn{ext} = - \zeroN q E(t) R \lt( n - n_0 \rt) \had_n \ha_n,
	\end{align}
models the effect of an electric field with strength $E(t)$ on the potential energy of a localised electron with charge $-q$. The potential energy due to $E(t)$ is set to zero at some arbitrary $n_0$, and $R = 4.5\tn{\ang}$ is the equilbrium lattice spacing.  Since the electron mass is several orders smaller than the mass of a peptide unit, we take a semi-classical approach where the phonon Hamiltonian, $\hHp$, is a classical one. In the harmonic approximation, the hydrogen bonds are modelled as Hookean springs with force constant $K$, and therefore $\hHp$ takes the form
	\bal
	\hHp = \zeroN \fr{P_n^2}{2M} + \zeroNmone M\O^2 ~\fr{\lt(U_{n+1} - U_{n}\rt)^2}{2}, \label{DSham_p}
	\end{align}
where $M = 1.774 \times 10^{-25}\tn{kg}$ is the average mass of a peptide unit in a membrane \ahelix~\cite{Langelaan2010}, and we have defined $\O := \sqrt{K/M}$. $U_n$ and $P_n$ are, respectively, the displacement and conjugate momentum of the $n\tsups{th}$ unit. Thus, the first and second sums in the expression for $\hHp$ represent, respectively, the kinetic and potential energies of the lattice. We take the value of $\O$ to be the natural angular frequency of slow phonons in an \ahelix, $\O = 5.5 \times 10^{12} \tn{s}^{-1}$ \cite{Brown1972,Chou1983,Chou1984}. To derive the interaction Hamiltonian, $\hHint$, Davydov and Scott assumed that the energy of an on-site excitation depends on lattice deformations in its vicinity. For us, the local deformations are 
	\bal
	S_n := U_{n+1} - U_n,
	\end{align} 
namely the amount by which the lengths of hydrogen bonds deviate from equilibrium. By Davydov and Scott's assumption, if we write the electron energy at site $n$ in a Taylor expansion, the first two terms are $J_0 + \chi G_n(S_n,S_{n-1})$, where $\chi$ is a constant, $G_n$ is a linear function, and $\abs{\chi G_n / J_0} \ll 1$. Then the interaction Hamiltonian is $\hHint = \zeroN \chi G_n \had_n \ha_n$. Davydov assumed that $S_n$ and $S_{n-1}$ have equal influence on local excitation energies \cite{Davydov1982}, so $G_n=\lt(S_n+S_{n-1}\rt)/2$, and $\hHint$ is
	\bal
	\hHint^{\tn{Dav}} = (\chi/2)[(U_1-U_0) \had_0 \ha_0 + \oneNmone (U_{n+1}-U_{n-1}) \had_n \ha_n + (U_N-U_{N-1}) \had_N \ha_N]. \label{HintDav}
	\end{align}
Davydov's model is therefore spatially symmetric, since $\had_n \ha_n$ is coupled equally to $U_{n+1}$ and to $U_{n-1}$. Scott modified Davydov's model by opting for the antisymmetric $G_n(S_n,S_{n-1})=S_n$ instead \cite{Scott1992}. The reason is, while both authors let an intra-peptide C=O oscillator take the role of the exciton, Davydov neglected the internal geometry of the peptide units, but Scott pointed out that every unit has its C=O pair immediately adjacent to the next hydrogen bond in the chain. This leads Scott to assume that $\had_n \ha_n$ is coupled to, without loss of generality, $S_n$, and not $S_{n-1}$. Scott therefore had 
	\bal
	\hHint^{\tn{Sco}} = \zeroNmone \chi(U_{n+1}-U_n) \had_n \ha_n. \label{HintSco}
	\end{align}
Since we are modelling an electron as opposed to an intra-peptide oscillator, we cannot use Scott's argument to justify assuming that $\had_n \ha_n$ is coupled to $U_{n+1}$ and not to $U_{n-1}$. Nor should we assume that on-site energies are affected equally by deformations on both sides, as Davydov did. We therefore propose $G(S_n,S_{n-1}) = \chir S_n + \chil S_{n-1}$, taking without loss of generality $\chir > 0$ and $0 \leq \chil \leq \chir$. Then, 
	\bal
	\hHint = ~& \chir (U_1-U_0) \had_0 \ha_0 + \chil ( U_N - U_{N-1} ) \had_N \ha_N \nonumber \\
	&+ \oneNmone [ \chir (U_{n+1}-U_n) + \chil (U_n - U_{n-1}) ] \had_n \ha_n.
	\end{align}
By defining
	\bal
	\chi := \chir + \chil, \quad \b = \fr{\chir-\chil}{\chir+\chil} \in \lt[ 0, 1 \rt], \label{anisotropy}
	\end{align}
we can write
	\bal
	\hHint = ~& \fr{\chi}{2}(1+\b) (U_1 - U_0) \had_0 \ha_0 + \fr{\chi}{2}(1-\b) (U_N - U_{N-1}) \had_N \ha_N \nonumber \\
	&+ \oneNmone \fr{\chi}{2} \lt[ \lt(U_{n+1} - U_{n-1}\rt) + \b \lt( U_{n+1} + U_{n-1} - 2U_n \rt) \rt] \had_n \ha_n. \label{DSham_int}
	\end{align}
We treat $\chi$ as an adjustable parameter. Setting $\b = 0$ ($\chil = \chir$) gives us the symmetric model of Davydov as per \cref{HintDav}, whilst setting $\b = 1$ ($\chil = 0$) produces the antisymmetric model of Scott as per \cref{HintSco}. The larger $\b$ is, the less spatial symmetry our model possesses. Indeed, for $\b \in [0,1)$, the ratio of $n$-$(n+1)$ coupling strength to $n$-$(n-1)$ coupling strength is given by $\chir/\chil = (1+\b) / (1-\b)$, and this ratio is strictly increasing with $\b$. 

We write the electronic state of the system as a linear superposition of local excitations \cite{Brizhik2008}, 
	\bal
	\ktPst = \sum_{n=0}^N \a_n(t) \had_n \ktvac, \label{polaronstate}
	\end{align}
where $\ktvac$ is the vacuum state, and $\a_n \in \mbb{C}$ is the probability amplitude for an electron localised at the $n$\tsups{th} site, subject to the normalisation condition,
	\bal
	\sum_{n=0}^N \abs{\a_n}^2 = 1. \label{discrete_norm_cond1}
	\end{align}
We proceed to derive dynamical equations for $\a_n$ and $U_n$. By equating coefficients of $\had_n \ktvac$ on both sides of the \SE, $i \hb ~\d \ktPs / \d t = (\hHe + \hHint + \hH_\tn{ext}) \ktPs$, we obtain
	\bal
	i\hb\fr{\d \a_n}{\d t} = &\lt[ J_0 + \fr{\chi}{2} \lt(S_n + S_{n-1} \rt) + \fr{\chi}{2} \b \lt( S_n - S_{n-1} \rt) \rt] \a_n - J_1 \lt(\a_{n+1} + \a_{n-1} \rt) \nonumber \\
	&- e E(t) R \lt( n-n_0 \rt) \a_n. \label{statican} 
	\end{align}
We have defined 
	\bal
	S_{-1} = S_N = 0, \quad \a_{-1} = \a_{N+1} = 0, \label{convenience1}
	\end{align}
so that \cref{statican} holds for all $n$ including the boundary terms ($n=0,N$). Equations for $U_n$ are derived from classical Hamilton equations, $\d U_n / \d t = \p H_\tn{cla} / \p P_n$ and $\d P_n / \d t = - \p H_\tn{cla} / \p U_n$, where $H_\tn{cla} := \brk{\Ps | (\hHp + \hHint) | \Ps}$. These equations are
	\bal
	M\fr{\d^2 U_n}{\d t^2} = &\lt(S_n - S_{n-1}\rt) + \fr{\chi}{2} \lt[\lt( \anponesq + \ansq \rt) - \lt( \ansq + \anmonesq \rt) \rt] \nonumber \\
	&- \fr{\chi}{2}\b \lt[ \lt( \anponesq - \ansq \rt) - \lt( \ansq - \anmonesq \rt)  \rt]. \label{staticUn}
	\end{align}
In order that \cref{staticUn} holds at the boundaries, we have set
	\bal
	\a_0 = \a_N = 0. \label{convenience2}
	\end{align}
This boundary condition is justified because we expect the probability distribution $\abs{\a_n}^2$ to be highly localised with half-width of $\mcal{O}(1)$, and because we will be working with large lattices with $N \gg 1$. We also impose the following boundary condition on $U_n$, representing a peptide chain which is fixed at one end.
	\bal
	U_0 = \fr{\d U_0}{\d t} = 0. \label{Un_boundary}
	\end{align}
Next, we introduce the gauge transformation, 
	\bal
	\a_n (t) = \ps_n (t) \exp \lt[ -\fr{it}{\hb} \lt( J_0 - 2J_1 \rt) \rt], \label{gauge}
	\end{align}
which sets $J_0=2$ in \cref{statican}. Physically this represents a shift in the arbitrary reference value from which energy is measured. Combining the $2\a_n$ term with the $J_1$ term in \cref{statican}, we obtain the discrete Laplacian, $-J_1 \lt( \a_{n+1} + \a_{n-1} - 2 \a_n\rt)$. Meanwhile, to account for the interaction between the peptide chain and its environment, we need to add \emph{Langevin terms} to the r.h.s. of \cref{staticUn} \cite{Brizhik2010,Brizhik2014,Lemons1997,Schlick2010}. They are, a damping term describing energy dissipation due to friction, $-\Gm ~\d U_n / \d t$, where $\Gm$ is the viscous damping coefficient; and a stochastic term $F_n(t)$, describing random forces due to thermal fluctuations. Specifically, $F_n(t)$ is normally-distributed with zero mean and correlation function $\brk{F_m (t)F_n (t')} = 2 \Gm k_B \Theta \de_{m,n} \de(t-t')$, where $k_B$ is the Boltzmann constant and $\Theta$ is the temperature of the environment. 

Scaling time by $\O^{-1}$ and length by $R$ gives us the following non-dimensionalised dynamical equations for $\ps_n$ and $u_n := U_n / R$, for $n=0,1,\dots,N$.
	\bsubs \label{dimlesseqns}
	\bal
	i \dotpsn &= \s \lt[ \lt(s_n + s_{n-1}\rt) + \b \lt( s_n - s_{n-1} \rt) \rt] \ps_n - \rh \lt( \ps_{n+1} + \ps_{n-1} - 2\ps_n \rt) - \eps(\t) (n - n_0) \ps_n, \label{dimlessphn} \\
	\ddotu_n &= \lt(s_n - s_{n-1}\rt) + \de \lt[ \lt(c_n - c_{n-1}\rt) - \b \lt( g_n - g_{n-1} \rt) \rt] - \gm \dotu_n + f_n(\t), \label{dimlessun}
	\end{align}
	\esubs
where we have defined
	\bal
	s_n := u_{n+1}-u_n, \quad g_n := \psnponesq - \psnsq, \quad c_n := \psnponesq + \psnsq,
	\end{align}
and where the overdot denotes differentiation with respect to dimensionless time $\t$, and
	\bal
	\rh = \fr{J_1}{\hb \O}, \quad \s = \fr{R \chi}{2\hb \O}, \quad \eps = \fr{qER}{\hb \O}, \quad \de = \fr{\chi}{2MR\O^2}, \quad \gm = \fr{\Gm}{M \O}, \quad f_n = \fr{F_n}{MR\O^2}. \label{dimlessparams}
	\end{align}
\Cref{dimlesseqns} holds subject to the boundary conditions (\ref{convenience1}), (\ref{convenience2}) and (\ref{Un_boundary}), as well as the normalisation condition,
	\bal
	\sum_{n=0}^N \abs{\ps_n}^2 = 1. \label{discrete_norm_cond2}
	\end{align}
It is easily verifiable that setting $\b = 0$ and $\b = 1$ in \cref{dimlesseqns} produces Davydov's and Scott's dynamical equations, respectively \cite{Davydov1982,Scott1992}. $\rh$ is known as an \emph{adiabaticity parameter}, as it is the ratio of the characteristic time scale of phonon vibrations to that of electronic phase variations \cite{Hennig2001}. We fix $J_1$, following \cite{Brizhik2008,Brizhik2010,Brizhik2014} (which used a different scaling), at $\rh = 2.1$. Moreover, since $M,R$ and $\O$ are fixed, the ratio $\s / \de = 1880$ is constant. The range of $\de$ which we consider throughout this study correspond to $\chi \sim \mcal{O}(10^{-11})$ Newtons, agreeing with \cite{Scott1992}. Finally, we take $\gm = 0.05$, agreeing with \cite{Brizhik2008,Brizhik2010,Brizhik2014} up to different scaling factors.


	\section{Stationary polaron solutions} \label{section3}

We derive stationary polaron solutions to \cref{dimlesseqns}, subject to zero electric field ($\eps = 0$) and zero temperature ($f_n = 0$). We consider analytical and numerical methods separately, and compare the results.



	\subsection{Analytical results} \label{subsection_anal}

When $f_n = 0$ and $\dotu_n = \ddotu_n = 0$, \cref{dimlessun} becomes
	\bal
	s_n - s_{n-1} = \de \lt[ \b \lt(g_n - g_{n-1} \rt) - \lt( c_n - c_{n-1} \rt) \rt],
	\end{align}
which holds if
	\bal
	s_n \equiv u_{n+1} - u_n = \de \lt( \b g_n - c_n \rt) = \de \lt[ \lt( \b - 1 \rt) \psnponesq - \lt( \b + 1 \rt) \psnsq \rt]. \label{steady_un}
	\end{align}
Putting \cref{steady_un} into \cref{dimlessphn} and requiring $\eps = 0$ gives us
	\bal
	i \dotpsn = &- \s \de \lt[ \lt( 1 - \b^2 \rt) \psnponesq + \lt( 1 - \b^2 \rt) \psnmonesq + 2 \lt( 1 + \b^2 \rt) \psnsq \rt] \ps_n \nonumber \\
	&- \rh \lt( \ps_{n+1} + \ps_{n-1} - 2\ps_n \rt) . \label{nls_discrete_comp}
	\end{align}
Defining
	\bal
	\De \ps_n &:= \ps_{n+1} + \ps_{n-1} - 2\ps_n, \qquad \De \psnsq := \psnponesq + \psnmonesq - 2\psnsq, \label{defn_disc_laplace} \\
	\lm &:= \fr{4\s \de}{\rh} \equiv \fr{\chi^2}{M \O^2 J_1}, \qquad \eta := \fr{\s \de}{\rh} \lt(1-\b^2\rt) \equiv \fr{\lm}{4} \lt(1-\b^2\rt), \label{defn_eta}
	\end{align}
we can rewrite \cref{nls_discrete_comp} as
	\bal
	i \rh^{-1} \dotpsn + \De \ps_n + \lm \psnsq \ps_n + \eta \De \psnsq \ps_n = 0 .  \label{nls_discrete}
	\end{align}
We note that, since $M,\O$ and $J_1$ are all fixed, the parameter $\lm$ inherits the adjustability of $\chi$. Now, in a stationary state, the time dependence of $\ps_n$ can be at most a variation of the phase factor. Following \cite{Davydov1982}, we consider the ansatz
	\bal
	\ps_n(\t) =  \lt. \exp \lt(i\rh H_0 \t + i k \xi \rt) \ph (\xi) \rt\vert_{\xi = (n-N/2)R}, \label{psn_ansatz}
	\end{align}
where $\xi$ is a real, continuous variable in the domain $[-NR/2, NR/2]$, $\ph$ is a real, twice-differentiable function, and $H_0$ and $k$ are constants. In particular, $H_0$ is an energy eigenvalue, in the sense that
	\bal
	i\rh^{-1} \dotpsn = -H_0\ps_n. \label{irhodotpsn}
	\end{align}
In the limit $N \gg 1$, $R$ becomes small compared to the domain size, which enables us to invoke the continuum approximation,
	\bal
	\ps_{n\pm 1} = \lt. \exp \lt(i\rh H_0 \t + i k \lt( \xi \pm R \rt) \rt) \lt( \ph(\xi) \pm R \ph'(\xi) + \fr{R^2}{2} \ph''(\xi) + \mcal{O}(R^3) \rt) \rt\vert_{\xi = (n-N/2)R}, \label{cont_app_ps} 
	\end{align}
implying
	\bal
	\abs{\ps_{n\pm 1}}^2 = \lt. \ph(\xi)^2 \pm 2R \ph(\xi) \ph'(\xi) + R^2 \ph(\xi) \ph''(\xi) + R^2 \lt( \ph'(\xi) \rt)^2 + \mcal{O}(R^3) \rt\vert_{\xi = (n-N/2)R}. \label{cont_app_pssq}
	\end{align}
Putting \cref{psn_ansatz,irhodotpsn,cont_app_ps,cont_app_pssq} into \cref{nls_discrete}, then dividing the result by $\exp \lt(i\rh H_0 \t + i k \xi \rt)$ and retaining terms up to $\mcal{O}(R^2)$, we obtain
	\bal
	0 = -H_0 \ph(\xi) &+ \lt[ \cos (k R) \lt( 2\ph(\xi) + R^2 \ph''(\xi) \rt) - 2\ph(\xi) + i \sin (k R) \lt( 2R \ph'(\xi) \rt) \rt] \nonumber \\
	&+ \lt. \lm \ph(\xi)^3 + \eta R^2 \lt[ 2 \ph(\xi) \ph''(\xi) + 2 \lt( \ph'(\xi) \rt)^2 \rt] \rt\vert_{\xi = (n-N/2)R}. \label{complex_eqn}
	\end{align}
The last term in \cref{complex_eqn} is equivalent to $\eta R^2 ~\d^2 (\ph(\xi))^2 / \d \xi^2$. Equating imaginary parts of \cref{complex_eqn} gives us $k = 0$. After the scaling $x:= \xi / R$, the real part of \cref{complex_eqn} becomes
	\bal
	-H_0 \ph + \ph_{xx} + \lm \ph^3 + \eta (\ph^2)_{xx} \ph = 0 \quad \tn{when}~x = n-N/2. \label{eqn}
	\end{align}
The subscript $x$ denotes differentiation with respect to $x$. We seek $\ph(x)$ which satisfies \cref{eqn} \emph{for all} $x$, not just when $x = n-N/2$. Then, from such a $\ph(x)$ we will be able to recover the discrete solution $\ps_n(\t)$ via $\xi = x R$ and \cref{psn_ansatz}. Further to being globally defined, we require that $\ph(x)$ has vanishing derivatives at infinity, and satisfies the normalisation condition,
	\bal
	\int_{-\infty}^\infty \ph(x)^2 ~\d x = 1. \label{norm_cond}
	\end{align}
If $\eta = 0$ (i.e. $\b = 1$), then \cref{eqn} reduces to the nonlinear \SE~with a cubic nonlinearity, which has a well-known solution satisfying all the above constraints \cite{Scott1992},
	\bal
	 H_0 = \lm^2 / 16, \quad \ph(x) = \pm \sqrt{\fr{\lm}{8}} ~\sech \fr{\lm x}{4} \quad \tn{for all}~x. \label{scott_soln}
	\end{align}
Consider $\eta > 0$ (i.e. $\b < 1$). Since $(\ph^2)_{xx} \equiv 2\ph \ph_{xx} + 2(\ph_x)^2$, we rewrite \cref{eqn} as
	\bal
	& -H_0 \ph + \ph_{xx} \lt( 1 + 2 \eta \ph^2 \rt) + \lm \ph^3 + 2 \eta \lt( \ph_x \rt)^2 \ph = 0. \label{eqn_re}
	\end{align}
We dedicate the remainder of this section to analysing \cref{eqn_re}. It is an autonomous equation for $\ph(x)$, which allows us to define $h(\ph) := \ph_x$, and write
	\bal
	\ph_{xx} \equiv \fr{\d (\ph_x)}{\d \ph} \ph_x = h h_\ph. \label{phphxsq}
	\end{align}
We define $y(\ph) := h^2$, so that $y_\ph = 2hh_\ph$, and multiply \cref{eqn_re} by 2 to obtain
	\bal
	\lt(1+2\eta\ph^2\rt) y_\ph + 4 \eta \ph y = 2 H_0 \ph - 2 \lm \ph^3. \label{linear}
	\end{align}
The l.h.s. of \cref{linear} is the total derivative of $(1+2\eta\ph^2) y$ with respect to $\ph$. We therefore have
	\bal
	y(\ph) = \fr{\int \lt( 2 H_0 \ph - 2 \lm \ph^3 \rt) \d \ph}{1 + 2 \eta \ph^2} = \fr{H_0 \ph^2 - \lm \ph^4 / 2 + C}{1 + 2 \eta \ph^2}. \label{soln_y}
	\end{align}
The integration constant $C$ is determined by considering the limit $x \rrr \infty$, in which $\ph^2 \rrr 0$ and $y \equiv (\ph_x)^2 \rrr 0$. We therefore have $C = 0$. Now we note that, if $H_0 \le 0$, then the r.h.s. of \cref{soln_y} is negative whenever $\ph \neq 0$, so it cannot equal the l.h.s. which is $(\ph_x)^2$. Thus, if $H_0 \le 0$ then the only $\ph(x)$ satisfying \cref{soln_y} is identically zero. We therefore require $H_0 > 0$. Multiplying \cref{soln_y} by $4\ph^2$, we obtain
	\bal
	\lt( 2 \ph \ph_x \rt)^2 = \fr{4 H_0 \ph^4 - 2 \lm \ph^6}{1 + 2 \eta \ph^2}. \label{ph_x_over_ph_sq}
	\end{align}
We then define $\Ph := \ph^2$, and \cref{ph_x_over_ph_sq} becomes
	\bal
	\lt(\Ph_x\rt)^2 =\fr{4 H_0 \Ph^2 - 2 \lm \Ph^3}{1 + 2 \eta \Ph}. \label{eqn_Ph}
	\end{align}
If \cref{eqn_Ph} has a solution which is globally non-negative and twice-differentiable, has vanishing derivatives at infinity, and satisfies
	\bal
	\int_{-\infty}^\infty \Ph(x) ~\d x = 1, \label{norm_cond2}
	\end{align}
then we claim that $\Ph(x)$ must attain its global upper bound of $2H_0/\lm$ at some finite $x$, and that every local maximum of $\Ph$ must also be a global maximum. The proof of this claim is as follows. Since $\Ph$ cannot be identically zero, and since $\lim_{x\rrr\pm\infty}\Ph(x) = 0$, $\Ph(x)$ must have at least one turning point, at some finite $x$ and non-zero $\Ph$. But we observe from \cref{eqn_Ph} that $\Ph_x$ vanishes if and only if $\Ph = 0$ or $ 2H_0 / \lm $. Therefore, $\Ph(x)$ must attain its global upper bound of $2H_0/\lm$ at least once, and no other local maximum value is possible. This concludes the proof. We further propose that wherever $\Ph(x)$ attains its maximum value, say at $x=\xmax$, the second derivative $\Ph_{xx}$ does not vanish there. The proof is as follows. On the one hand, we have $\Ph_{xx} \equiv 2\ph \ph_{xx} + 2(\ph_x)^2$; when $x = \xmax$, we also have $\ph_x = \Ph_x / (2\ph) = 0$, therefore $\Ph_{xx} = 2\ph \ph_{xx}$. On the other hand, \cref{eqn} is equivalent not only to \cref{eqn_re} but also to $\eta \ph \Ph_{xx} = H_0 \ph - \ph_{xx} - \lm \ph^3$. It follows that, at $x = \xmax$, we have $(1 / (2\ph) + \eta \ph) \Ph_{xx} = H_0 \ph - \lm \ph^3$, and therefore $\Ph_{xx} = (2H_0 \Ph - 2\lm \Ph^2) / (1 + 2\eta \Ph)$. Since $\Ph(\xmax) = 2H_0 / \lm$, it follows that $\Ph_{xx}(\xmax) = -4H_0^2 / (\lm + 4\eta H_0) < 0$, as required. A corollary of this proposition is that there must exist some neighbourhood of $\xmax$ containing no maxima of $\Ph(x)$ other than $\xmax$ itself. Without loss of generality, let $\xmax=0$. Suppose the corollary is false, so that every neighbourhood of 0 contains some non-zero $x$ which is a maximum of $\Ph(x)$. Then, there must exist some sequence $x_n$ approaching 0 such that $\Ph(x)$ has a maximum at every $x_n$, with $\Ph(x_n) = \Ph(0)$. But this leads to a contradiction. Indeed, for every $x_n$ we have the Taylor expansion $\Ph(x_n) = \Ph(0) + \Ph_{xx}(0) x_n^2 / 2 + \mcal{O} (x_n^3)$, where the first derivative is absent because $\Ph(x)$ has a maximum at 0. It then follows that $\Ph_{xx}(0) = \lim_{n \rrr \infty} 2 (\Ph(x_n) - \Ph(0)) / x_n^2 = 0$, which contradicts the previous proposition. Therefore the corollary is proven. Since 0 has a maxima-free neighbourhood, we say that $\Ph(x)$ has an \emph{isolated maximum} at 0.

We note that we can indeed \emph{require} that $x=0$ is a maximum of $\Ph(x)$, because \cref{eqn_Ph} is translationally invariant: if $\Ph(x)$ is a solution then so is $\Ph(x-c)$ for any constant $c$. We exploit this invariance, requiring that $\Ph(x)$ satisfies
	\bal
	\Ph_0 := \Ph(0) = 2H_0/\lm. \label{ana_init_cond}
	\end{align}
Now we claim that there exists $b > 0$, which may be infinite, such that $\lim_{x \rrr b} \Ph(x) = 0$, and $\Ph(x) \neq 0$ for all $x \in (0,b)$. The proof of this claim is as follows. If $\Ph(x) \neq 0$ for all $x \in (0,\infty)$, then we are done. If $\Ph(x) = 0$ for some $x \in (0,\infty)$, then the set $\{ x \in (0,\infty) : \Ph(x) = 0 \}$ must have a minimum. If it does not, then there would be a sequence $x_n > 0$ such that, as $n \rrr \infty$, $x_n \rrr 0$ and $\Ph(x_n) \rrr 0$; but this would contradict the continuity of $\Ph(x)$ at $x=0$. Thus, letting $b$ equal the least positive zero of $\Ph(x)$, then we are done.

Next, we propose that no other solution on $[0,b)$ exists, and the proof is as follows. If $\Ph(x)$ has any maxima in $(0,\infty)$, then the set $\mcal{M} := \{x \in (0,\infty) : x~ \tn{is a maximum of}~ \Ph(x)\}$ must have a minimum, because otherwise we would have a contradiction to the fact that $x = 0$ is an \emph{isolated} maximum of $\Ph(x)$. Let $x_1 = \min \mcal{M}$, and suppose $x_1 < b$. Since $\Ph(x_1) = \Ph(0)$, and since $\Ph(x)$ has no maximum in the interval $(0,x_1)$, and since $\Ph(x)$ is continuous, it must attain its minimum value at some point $b' \in (0,x_1)$. But that implies $\Ph(b') = 0$, where $b' < x_1 < b$, contradicting the fact that $\Ph(x) \neq 0$ for all $x \in (0,b)$. Therefore, we must have $x_1 > b$. Since $\Ph_x$ vanishes only at maxima and minima, it follows that $\Ph_x$ is non-vanishing on $(0,b)$, and therefore $\Ph(x)$ is \emph{strictly} decreasing on $[0,b)$. That is, \emph{any} solution to \cref{eqn_Ph} on $[0,b)$ satisfying all the aforementioned contraints must also satisfy
	\bal
	\Ph_x = - g(\Ph) := - \sqrt{\fr{4 H_0 \Ph^2 - 2 \lm \Ph^3}{1 + 2 \eta \Ph}} = -2\sqrt{H_0} \Ph \sqrt{\fr{1 - \Ph / \Ph_0}{1 + 2 \eta \Ph}}, \quad 0 \le x < b,~ \Ph_0 \ge \Ph > 0. \label{negroot}
	\end{align}
On any closed interval $[\Ph_1,\Ph_2] \subset (0,\Ph_0)$, the function $g(\Ph)$ is continuous and non-zero, so the reciprocal function $1/g(\Ph)$ is continuous and bounded, and therefore Riemann integrable. But $g(\Ph)$ approaches 0 as $\Ph \rrr \Ph_0$, meaning $1/g(\Ph)$ becomes unbounded. Thus, integration of $1/g(\Ph)$ on the interval $[\Ph_1,\Ph_0]$ is not trivial. Luckily $\Ph = \Ph_0$ is an \emph{integrable singularity} of the function $1/g(\Ph)$, because the Puiseux series of $1/g(\Ph)$ about $\Ph_0$ is $\mcal{O} ((\Ph - \Ph_0)^{-1/2})$. Therefore, for any $\Ph_1 \in (0,\Ph_0]$, \cref{negroot} is equivalent to 
	\bal
	\int_{\Ph_1}^{\Ph_0} \fr{1}{g(\Ph)} ~\d \Ph = x(\Ph_1) - x(\Ph_0) = x(\Ph_1), \label{inversesoln}
	\end{align}
which determines a unique $x(\Ph_1) \in [0,b)$. The l.h.s. of \cref{inversesoln} is a strictly decreasing function of $\Ph_1$, meaning $x(\Ph_1)$ has a unique inverse function which is also strictly decreasing, $\Ph_1(x)$, on the domain $x \in [0,b)$. But $\Ph(x)$ is an existing function satisfying \cref{negroot} for all $x \in [0,b)$. Therefore, we must have $\Ph_1(x) = \Ph(x)$ for all $x \in [0,b)$, and the uniquess of $\Ph(x)$ follows.

In summary, we have so far established the following. If \cref{eqn_Ph} has a solution which is globally non-negative and twice-differentiable, has vanishing derivatives at infinity, and has the property that its integral over $\mbb{R}$ is 1, then \cref{eqn_Ph} has a solution, say $\Ph(x)$, which satisfies all the above constraints as well as the condition (\ref{ana_init_cond}), and there exists some $b > 0$ which may be infinite such that $\Ph(x)$ is strictly decreasing on $[0,b)$, and $\lim_{x\rrr b^-} \Ph(x) = 0$, and $\Ph(x)$ is the unique solution on $[0,b)$. Moreover, using exactly the same arguments as above, it can also be shown that there exists some $a < 0$ which may be infinite such that $\Ph(x)$ is strictly increasing on $(a,0]$, and $\lim_{x \rrr a^+} \Ph(x) = 0$, and $\Ph(x)$ is the unique solution on $(a,0]$. On the \emph{interval of uniqueness}, $(a,b)$, $\Ph_x$ is given by
	\bal
	\Ph_x = G(x,\Ph) := -\sgn(x) g(\Ph), \label{Ph_x_allx}
	\end{align}
where $g$ is defined by \cref{negroot}, and $\sgn$ is the sign function. 

Now we describe a method which, given $\lm$ and $\eta$, determines the unique $\Ph(x)$ on $(a,b)$, and also determines $a,b,H_0$ in the process. Indeed we will show that for any $\lm$ and $\eta$, the interval of uniqueness for $\Ph(x)$ must be $(a,b) = \mbb{R}$. The fact that $(a,b) = \mbb{R}$ shall have the following subtle consequence. Note that the derivation of \cref{eqn_Ph} involved a multiplication by $\Ph \equiv \ph^2$. Thus, the deduction from \cref{eqn_Ph} back to \cref{eqn_re} holds \emph{on the condition that} $\Ph \neq 0$. Since $a$ and $b$ are the smallest (in magnitude) zeros of $\Ph(x)$, we see that the equivalence between \cref{eqn_Ph,eqn_re} breaks down outside the interval $(a,b)$. That is to say, \cref{eqn_Ph,eqn_re} are equivalent globally if and only if $(a,b) = \mbb{R}$.

The method is as follows. For $x \in [0,b)$, consider the coordinate transformation,
	\bal
	Z(\Ph) := \arsech \lt(Y(\Ph)\rt), \quad \tn{where~} Y(\Ph) := \sqrt{\fr{\Ph}{\Ph_0}} = \sqrt{\fr{\lm \Ph}{2 H_0}}. \label{defn_Z}
	\end{align}
$\Ph(x)$ is a bijection from $[0,b)$ to $(0,\Ph_0]$, $Y(\Ph)$ is a bijection from $(0,\Ph_0]$ to $(0,1]$, and the inverse sech function, arsech, is a bijection from $(0,1]$ to $[0,\infty)$. Therefore, all the coordinate transformations are invertible. For $x \in (a,0]$, we consider exactly the same transformations as \cref{defn_Z}.
Differentiating $Z$ with respect to $x$ we find, for all $x \in (a,b)$,
	\bal
	Z_x &= Z_Y \cdot Y_\Ph \cdot \Ph_x \nonumber \\
	&= \fr{-1}{Y \sqrt{1-Y^2}} \cdot  \fr{1}{2\sqrt{\Ph_0 \Ph}} \cdot \lt( - \sgn(x) g(\Ph) \rt) \nonumber \\
	&= \fr{\sgn(x)}{2 \Ph \sqrt{1 - \Ph / \Ph_0}} \cdot g(\Ph) \nonumber \\
	&= \fr{\sgn(x)\sqrt{H_0}}{\sqrt{1 + 2\eta \Ph}}, \label{Z_x_1}
	\end{align}
where we have used definition (\ref{negroot}) of $g(\Ph)$. Moreover, by definition we have $Y = \sech Z$, and it follows that $2 \eta \Ph = 2 \eta \Ph_0 Y^2 = 2 \eta (2H_0 / \lm) \sech^2 Z$. Defining 
	\bal
	\n := 4 \eta H_0 / \lm, \label{defn_nu}
	\end{align}
we rewrite \cref{Z_x_1} as
	\bal
	Z_x = \fr{\sgn(x)\sqrt{H_0}}{\sqrt{1 + \nu~ \sech^2 Z}}. \label{Z_x}
	\end{align}
Due to \cref{defn_Z}, we have $Z(x=0) = 0$. We can therefore solve \cref{Z_x} as follows.
	\bal
	\sgn(x) \sqrt{H_0} \int_{0}^{x} \d \tilde{x} = \int_{0}^{Z} \sqrt{1 + \nu~ \sech^2 \tld{Z}} ~\d \tld{Z}, \label{soln2_0}
	\end{align}
implying
	\bal
	\sgn(x) \sqrt{H_0} ~x = \arsinh \lt( \fr{\sinh Z}{\sqrt{1 + \n}} \rt) + \sqrt{\n} ~\arctan \lt( \fr{\sqrt{\n} ~\sinh Z}{\sqrt{\n + \cosh^2 Z}} \rt). \label{soln2}
	\end{align}
Now we can determine the values of $a$ and $b$. In the limit $Z \rrr +\infty$, the definition of the coordinate transformations, as per \cref{defn_Z}, dictates that we must have either $x \rrr a$ or $x \rrr b$. At the same time, \cref{soln2} dictates that we must have $x \rrr \pm \infty$, because the $\arctan$ function on the r.h.s. of \cref{soln2} is bounded, whilst the $\arsinh$ function diverges to $+\infty$. It therefore follows that $(a,b) = \mbb{R}$.

The next step is to rewrite \cref{soln2} as an expression for $x$ in terms of $\Ph$, so that we can invert the expression to find $\Ph(x)$ for $x \in \mbb{R}$. By definition (\ref{defn_Z}) we have $\cosh^2 Z = 1 / Y^2 = \Ph_0 / \Ph$, and it follows that $\sinh^2 Z = \cosh^2 Z - 1 = (\Ph_0 / \Ph) - 1$. Since $Z$ is by definition non-negative, we must take the positive square root, $\sinh Z = \sqrt{(\Ph_0 / \Ph) - 1}$. Then \cref{soln2} becomes
	\bal
	\sgn(x) \sqrt{H_0} ~x = \arsinh \sqrt{ \fr{1 - ( \Ph / \Ph_0 )}{\lt( 1+\n \rt) ( \Ph / \Ph_0 )} } + \sqrt{\n}~ \arctan \sqrt{ \fr{ \n \lt( 1 - (\Ph/\Ph_0) \rt) }{1 + \n (\Ph / \Ph_0)}} . \label{soln3}
	\end{align}
We claim that, given $\Ph_0 > 0$ and $x \in \mbb{R}$, \cref{soln3} uniquely determines a value of $\Ph > 0$. The proof is as follows. If $x = 0$, then immediately from \cref{soln3} we have $\Ph = \Ph_0$, and we are done. If $x \neq 0$, consider the function
	\bal
	\mcal{G} (\Ph) := \arsinh \sqrt{ \fr{1 - ( \Ph / \Ph_0 )}{\lt( 1+\n \rt) ( \Ph / \Ph_0 )} } + \sqrt{\n}~ \arctan \sqrt{ \fr{ \n \lt( 1 - (\Ph/\Ph_0) \rt) }{1 + \n (\Ph / \Ph_0)}} - \sgn(x) \sqrt{H_0} ~x, \label{defn_mcalG}
	\end{align}
where $x$ and $\Ph_0$ are parameters. Differentiating \cref{defn_mcalG} with respect to $\Ph$, we find
	\bal
	\fr{\d \mcal{G}}{\d \Ph} = -\fr{1}{2\Ph} \sqrt{\fr{ 1 + \n (\Ph / \Ph_0) }{ 1 - ( \Ph / \Ph_0 ) }} < 0 \quad ~\tn{for}~ \Ph > 0.
	\end{align}
This means $\mcal{G}$ is strictly decreasing for $\Ph > 0$. Since $\mcal{G}(\Ph) \rrr \infty$ in the limit $\Ph \rrr 0$, and $\mcal{G}(\Ph_0) = -\sgn(x) \sqrt{H_0} ~x < 0$, and $\mcal{G}$ is continuous, we must have $\mcal{G}$ vanishing at exactly one value of $\Ph \in (0,\Ph_0)$. This concludes the proof. Moreover, we observe that in \cref{soln3} the l.h.s. is invariant under $x \mapsto -x$. Thus, on $\mbb{R}$ we have $\Ph(-x) = \Ph(x)$.

In practice, given any $\Ph_0 > 0$ and $x \in \mbb{R}$, we can compute $\Ph(x)$ by locating the zero of $\mcal{G}(\Ph)$. However, the value of $\Ph_0$ cannot be freely chosen. Instead, it is determined by the normalisation condition (\ref{norm_cond2}) which, since $\Ph(x)$ is an even function on $\mbb{R}$, now reads
	\bal
	1 = 2 \int_0^\infty \Ph(x) ~\d x = 2 \int_{Z = 0}^\infty \fr{\Ph}{Z_x^+}~ \d Z, \label{rewrite_norm_cond}
	\end{align}
where the $Z_x^+$ is the positive-$x$ branch of $Z_x$, as per \cref{Z_x}. It then follows that
	\bal
	1 = 2 \int_0^\infty \fr{\Ph \sqrt{1 + \nu~ \sech^2 Z}}{\sqrt{H_0}} ~\d Z.
	\end{align}
Using $\Ph = \Ph_0 ~\sech^2 Z$, we deduce
	\bsubs \label{trans}
	\bal
	\fr{\sqrt{H_0}}{2\Ph_0} &= \int_0^\infty \sech^2 Z \sqrt{1 + \nu~ \sech^2 Z} ~ \d Z \label{trans0} \\
	&= \half + \fr{\lt(1+\n\rt) \arctan \sqrt{\n}}{2\sqrt{\n}}. \label{trans1}
	\end{align}
	\esubs
Multiplying \cref{trans1} by $2\sqrt{\n}$, replacing $H_0$ by $\lm \Ph_0 / 2$, and replacing $\n$ by $4 \eta H_0 / \lm = 2\eta \Ph_0$, we obtain the following transcendental equation for $\Ph_0$.
	\bal
	\sqrt{\lm \eta} = \sqrt{2 \eta \Ph_0} + \lt(1+ 2\eta \Ph_0 \rt) \arctan \sqrt{2\eta \Ph_0}. \label{trans_Ph0}
	\end{align}
To show that exactly one solution to \cref{trans_Ph0} exists, we consider the function
	\bal
	\mcal{F} (\Ph_0) := \sqrt{2 \eta \Ph_0} + \lt( 1 + 2 \eta \Ph_0 \rt) \arctan \sqrt{2\eta \Ph_0} - \sqrt{\lm \eta}.
	\end{align}
Differentiating $\mcal{F}(\Ph_0)$ with respect to $\Ph_0$, we find
	\bal
	\fr{\d \mcal{F}}{\d \Ph_0} = 2 \eta \lt( \fr{1}{\sqrt{2 \eta \Ph_0}} + \arctan \sqrt{ 2 \eta \Ph_0} \rt) > 0 \quad ~\tn{for}~\Ph_0 > 0.
	\end{align}
This means $\mcal{F}(\Ph_0)$ is strictly increasing for $\Ph_0 > 0$. Since $\lim_{\Ph_0 \rrr 0} \mcal{F}(\Ph_0) = -\sqrt{\lm \eta} < 0$, and $\mcal{F}(\Ph_0) \rrr \infty$ in the limit $\Ph_0 \rrr \infty$, and $\mcal{F}$ is continuous, we must have $\mcal{F}$ vanishing at exactly one value of $\Ph_0 > 0$. In practice, given parameters $\lm$ and $\eta$, we can compute $\Ph_0$ by locating the zero of $\mcal{F}(\Ph_0)$, and $\Ph_0$ uniquely determines the energy eigenvalue, $H_0 = \lm \Ph_0 / 2$. We can then feed the value $\Ph_0$ into \cref{soln3}, and then for every $x \in \mbb{R}$ we can find $\Ph(x)$ by means we have already described. In summary, given $\lm$ and $\eta$, \cref{soln3,trans_Ph0,ana_init_cond} together constitute an analytical solution to \cref{eqn_Ph}; and as we have already proven, it must be the unique global solution to \cref{eqn_Ph} which satisfies all the constraints we have imposed.

We note that if the parameter $\eta \rrr 0$, we should recover the solution to the nonlinear \SE, given by \cref{scott_soln}; and indeed we do. Firstly, in the limit $\eta \rrr 0$, we have $\n \rrr 0$, which means we cannot use \cref{trans_Ph0} to determine $\Ph_0$, because the derivation of \cref{trans_Ph0} involved a multipication by $\sqrt{\n}$. Instead, we must extract $\Ph_0$ from \cref{trans}. In the limit $\n \rrr 0$, \cref{trans0} is simply $\sqrt{\lm / (8\Ph_0)} = \int_0^\infty \sech^2 Z ~\d Z = 1$. It follows that $\Ph_0 = \lm/8$, agreeing with \cref{scott_soln}. Then \cref{ana_init_cond} determines the eigenvalue $H_0 = \lm \Ph_0 / 2 = \lm^2 / 16$, again agreeing with \cref{scott_soln}. Finally, when $\nu \rrr 0$, \cref{soln3} is simply 
	\bal
	\sgn(x) \sqrt{H_0} ~x = \arsinh \sqrt{ ( \Ph_0 / \Ph ) - 1 } ,
	\end{align}
which is equivalent to $\Ph_0 / \Ph = 1 + \sinh^2 (\sqrt{H_0}x) = \cosh^2 (\lm x / 4)$, so $\Ph = \Ph_0 ~\sech^2 (\lm x / 4)$, agreeing with \cref{scott_soln} once more.

The eigenvalue $H_0$ provides a link between $\Ph(x)$ and the binding energy of the stationary polaron. By \cref{polaronstate,gauge,psn_ansatz}, where $k = 0$ and $\xi  = xR$, the polaron state is written in terms of local excitations as $\ktPs = \sum_{n=0}^N \a_n \had_n \ktvac$, and in the limit $N \gg 1$, we have
	\bal
	\a_n = \ph(n - N/2) \exp \lt[ -\fr{it}{\hb} \lt( J_0 - 2J_1 - H_0 J_1 \rt) \rt], \quad \ph(x)^2 = \Ph(x). \label{stat64}
	\end{align}
Thus, the stationary $\ktPs$ solves $i \hb ~\d \ktPs / \d t = (\hHe + \hHint) \ktPs$ as well as satisfying $i \hb ~\d \ktPs / \d t = (J_0 - 2J_1 -H_0J_1) \ktPs$. By definition, the polaron's binding energy, $\Eb$, is its total internal energy measured with respect to $J_0$. In units of $J_1$, we have
	\bal
	\Eb := \fr{\brk{\Ps|\hHe+\hHp+\hHint|\Ps} - J_0}{J_1}. \label{bindingenergy1}
	\end{align}
An expression for $\brk{\hHp}$ in terms of $\Ph(x)$ can be found by using \cref{DSham_p,steady_un}. Since the polaron is stationary, the kinetic part of $\brk{\hHp}$ is zero, so we have
	\bal
	\fr{\brk{\hHp}}{J_1} =~& \fr{M \O^2 R^2}{2J_1} \zeroNmone \lt(u_{n+1} - u_{n}\rt)^2 \nonumber \\
	=~& \fr{\s \de}{2 \rh} \zeroNmone \lt[ \lt( \b - 1 \rt) \psnponesq - \lt( \b + 1 \rt) \psnsq \rt]^2 \nonumber \\
	=~& \fr{\lm}{8} \zeroNmone \lt[ \lt( \b - 1 \rt)^2 \abs{\ps_{n+1}}^4 + \lt( \b + 1 \rt)^2 \abs{\ps_n}^4  \rt] + \eta \zeroNmone \abs{\ps_{n+1}}^2  \abs{\ps_n}^2. \label{bindingenergy2}
	\end{align}
It then follows that 
	\bal
	\Eb =~& - 2 - H_0 + \fr{\brk{\hHp}}{J_1} \nonumber \\ 
	=~&- 2 - H_0 + \fr{\lm}{8} \zeroNmone \lt[ \lt( \b - 1 \rt)^2 \Ph(n+1-N/2)^2 + \lt( \b + 1 \rt)^2 \Ph(n-N/2)^2 \rt] \nonumber \\
	&+ \eta \zeroNmone \Ph(n+1-N/2) ~\Ph(n-N/2).  \label{bindingenergy3}
	\end{align}
We have made use of definition (\ref{defn_eta}) of $\lm,\eta$, the fact that $\abs{\ps_n} = \abs{\ph_n}$ for all $n$, as well as the fact that $\abs{\ph_n}^2$ is approximated by $\Ph(n-N/2)$. \Cref{fig_anal} shows how various aspects of the stationary polaron depends upon the \emph{symmetry parameter} $\b$, and the \emph{effective coupling parameter} $\lm$. Recall that the former is a measure of the spatial symmetry of the electron-phonon interaction, and the latter measures the strength of this interaction. These are the only two parameters that affect the stationary polaron's physical properties (as $\eta$ is merely a convenient combination of $\b$ and $\lm$).

\Cref{fig_anal1} shows how $\Ph_0$ and the \emph{half-width} of the polaron varies with $\b$ and $\lm$. We define the half-width as the distance between the two $x$-values at which $\Ph(x) = \Ph_0/2$, and it is a measure of how localised the polaron is. As one would expect, the half-width is negatively correlated with $\Ph_0$, which is the maximum height of $\Ph(x)$. The figure shows $\Ph_0$ increasing with $\lm$, and half-width decreasing with $\lm$, and the rate of change of each quantity is greater given larger values of $\b$. That is to say, the more spatially asymmetric the electron-lattice interaction is, the more influential $\lm$ is. The figure also has the following implication on the accuracy of $\Ph(x)$ as an approximation to the discrete stationary solution to \cref{dimlesseqns}. In a discrete solution, $\ps_n = \exp (i\rh H_0 \t) \ph_n$, the physical interpretation of $\abs{\ps_n}^2$ is the probability of the electron being localised at the $n$\tsups{th} lattice site. Therefore, the normalisation condition is defined in terms of a sum, $\zeroN \psnsq = 1$, and consequently we must have $\psnsq \le 1$ for all $n$. When a continuum solution $\Ph(x)$ is used to approximate the discrete one, we have the relation $\Ph_0 \equiv \max{\abs{\ps_n}^2}$. Thus, any continuum solution with $\Ph_0 > 1$ cannot be reliable as an approximant. When $\b = 1$, $\Ph_0$ exceeds 1 if $\lm$ is greater than 8, since $\Ph_0 = \lm/8$. On the other hand, when $\b = 0$, we computed $\Ph_0$ for $\lm$ up to 100, and $\Ph_0$ remains less than 0.6.

In \cref{fig_anal2} we see that $H_0$ increases with $\lm$ whilst the polaron's binding energy gains magnitude, meaning the larger $\lm$ is the more energy is required to break up the polaron. Once again, the larger $\b$ is, the more rapidly these quantities vary with $\lm$. We note that the thick (black) curve for $H_0$, corresponding to $\b = 1$, is exactly the graph of $H_0 = \lm^2/16$, as per \cref{scott_soln}. Comparing \cref{fig_anal1,fig_anal2}, we see that a polaron which is more strongly bound has a larger $\Ph_0$ and a smaller half-width, i.e., it is more localised.

	\bg{figure}[h!]
	\centering
	\bg{subfigure}[h]{0.49\textwidth}
	\captionsetup{width=0.9\textwidth}
	\includegraphics[width=\textwidth]{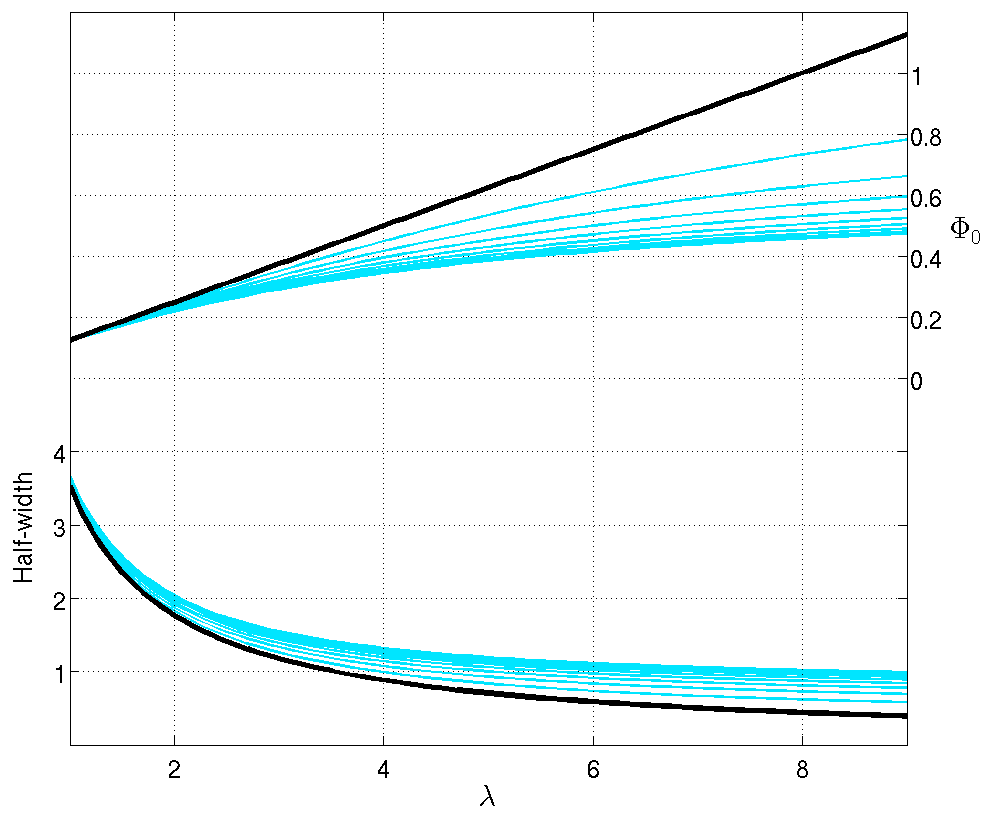} 
	\caption{ } \label{fig_anal1}
	\end{subfigure}
	\bg{subfigure}[h]{0.49\textwidth}
	\captionsetup{width=0.9\textwidth}
	\includegraphics[width=\textwidth]{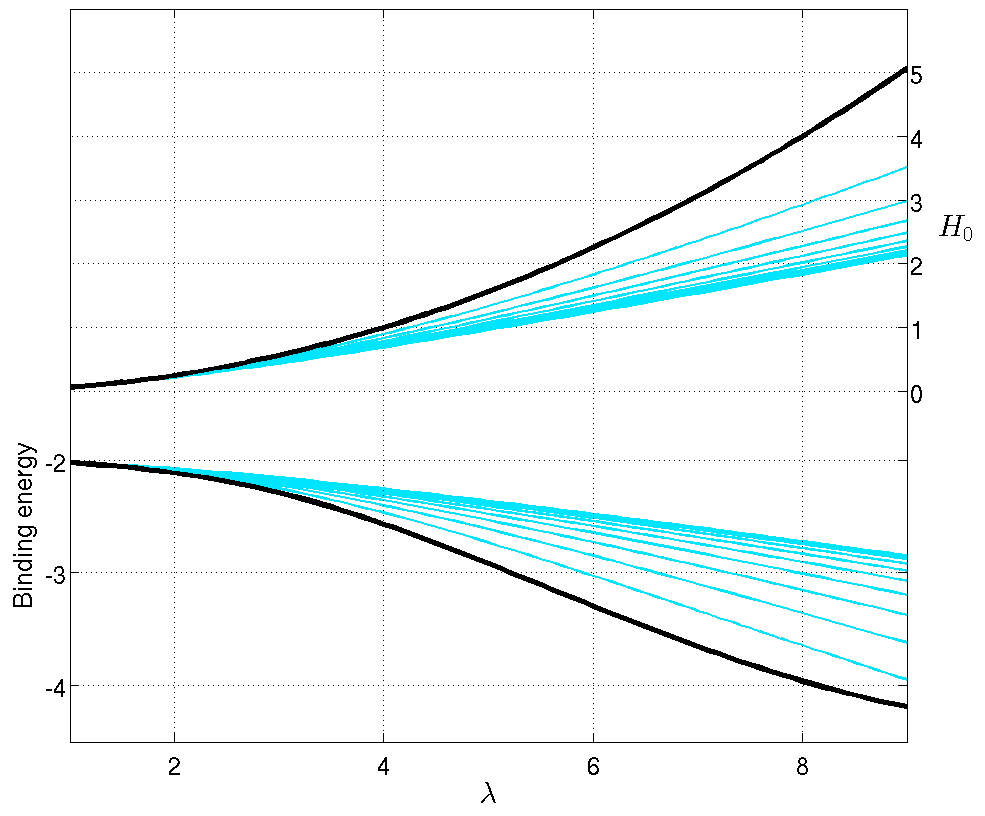} 
	\caption{ } \label{fig_anal2}
	\end{subfigure}
	\captionsetup{width=0.9\textwidth}
	\caption{(Colour online.) The height $\Ph_0$ of the analytical $\Ph(x)$ solution ((a), right axis), the polaron half-width ((a), left axis), the energy eigenvalue $H_0$ ((b), right axis), and the polaron binding energy ((b), left axis), according to analytical solutions $\Ph(x)$. The dependence of each quantity upon $\b$ and $\lm$ is represented by a family of curves. The thick (black) curve always corresponds to $\b = 1$, and as $\b$ decreases towards 0, the thin (blue) curves, corresponding to $\b = 0.9, 0.8, \dots, 0.1, 0$, become either steeper or shallower.} \label{fig_anal}
	\end{figure}
The gradient of curves in \cref{fig_anal2} vary with $\b$, and the variation is more pronounced when $\b$ is close to 1. This suggests that the system is highly sensitive to variations in $\b$ when $\b$ is large, but not so when $\b$ is small. Moreover, as $\lm$ decreases, curves corresponding to different values of $\b$ begin to converge; specifically this happens when $\lm \approx 1$. This suggests that when $\lm$ is small, the extent to which the electron-phonon interaction is spatially symmetric has little bearing on the physical properties of stationary polarons.


	\subsection{Numerical solutions} \label{subsection_num}

In this section we solve \cref{nls_discrete,irhodotpsn} directly, using a numerical scheme, but not without the help of analytical results from \cref{subsection_anal}. We then compare the resulting stationary polaron states with the ones we obtained via continuum approximation.

	\bg{figure}[h!]
	\centering
	\bg{subfigure}[h]{0.49\textwidth}
	\captionsetup{width=0.9\textwidth}
	\includegraphics[width=\textwidth]{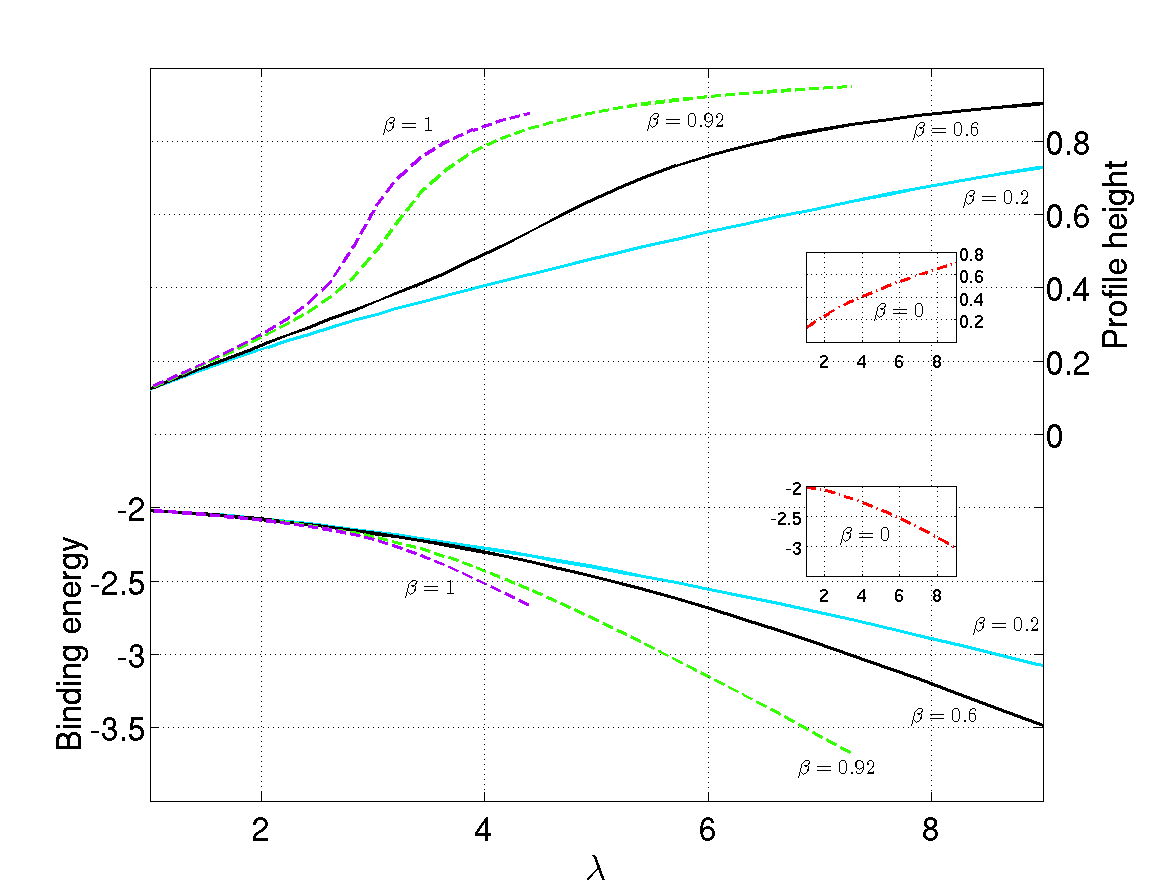}
	\caption{ } \label{fig_stat2}
	\end{subfigure}
	\bg{subfigure}[h]{0.49\textwidth}
	\captionsetup{width=0.9\textwidth}
	\includegraphics[width=\textwidth]{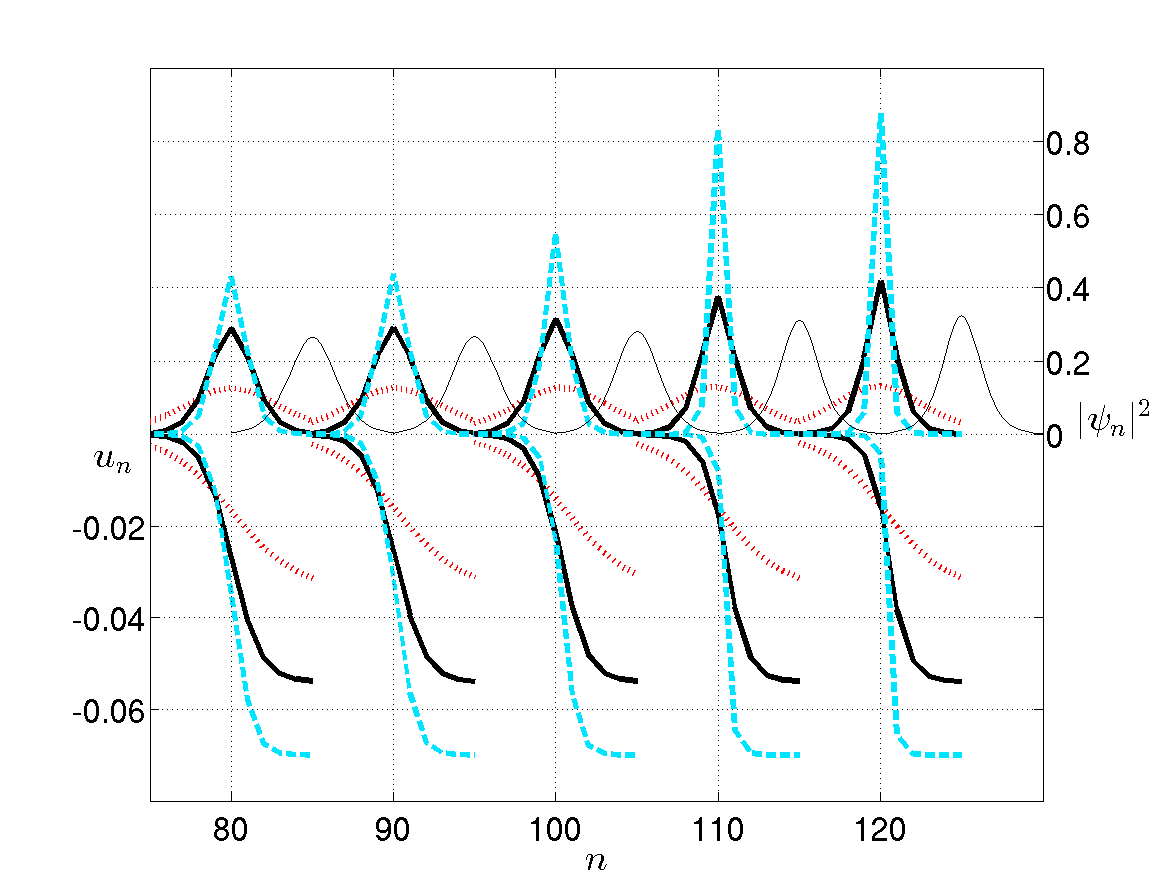}
	\caption{ } \label{fig_stat1}
	\end{subfigure}
	\captionsetup{width=0.86\textwidth}
	\caption{(Colour online.) (a) The maximum localisation probability $\maxpsnsq$ (right axis), and polaron binding energy (left axis), as functions of $\b$ and $\lm$. The curves for $\b = 0$ largely overlap those for $\b = 0.2$, so for practical reasons they are plotted on separate scales. For each value of $\b$, we are interested only in those $\lm$ for which $\maxpsnsq$ is not too close to the extremes, i.e. 0 and 1. \\ (b) Thick lines: numerical stationary solutions, $\psnsq$ (right axis), and the associated $u_n$ (left axis). Solutions are shifted along the $n$-axis, to avoid overlap. From left to right: $\b = 0,0.2,0.6,0.92,1$. Dotted (red) lines: $\lm = 1.0$. Solid (black) lines: $\lm = 2.6$. Dashed (blue) lines: $\lm = 4.4$. Thin lines: approximate solutions, by analytical methods of \cref{subsection_anal}. From left to right: $\b = 0,0.2,0.6,0.92,1$; all with $\lm = 2.6$.} \label{fig_stat}
	\end{figure}
Expanding \cref{nls_discrete} using the definitions of $\De \ps_n$ and $\De \psnsq$, we have
	\bal
	-H_0 \ps_n + \lt( \ps_{n+1} + \ps_{n-1} - 2\ps_n \rt) + \lm \psnsq \ps_n + \eta \lt( \psnponesq + \psnmonesq - 2\psnsq \rt)  = 0. \label{nls_discrete_re}
	\end{align}
Any solution $\ps_n$ to \cref{nls_discrete_re} is an attractor of the following map \cite{Kalosakas1998}.
	\bal
	\ps_n \mapsto \fr{\mcal{H} (\ps_n) }{\| \mcal{H} (\ps_n) \|}, \label{num_map}
	\end{align}
where 
	\bsubs
	\bal
	\mcal{H} (\ps_n) := \lt( \ps_{n+1} + \ps_{n-1} - 2\ps_n \rt) &+ \lm \psnsq \ps_n + \eta \lt( \psnponesq + \psnmonesq - 2\psnsq \rt) ,  \label{mcalHpsn}  \\
	\| \mcal{H}(\ps_n) \| &= \sqrt{\zeroN \mcal{H} ( \ps_n ) ^2}. 
	\end{align}
	\esubs
We take the approximate solution from \cref{subsection_anal} as initial guess, and repeatedly apply \cref{num_map} until we reach convergence. When converged, $\ps_n$ is the stationary solution to \cref{nls_discrete_re}, and $\| \mcal{H}(\ps_n) \|$ is equal to $H_0$. In practice, on a grid with $N = 200$, convergence is typically reached within $\mcal{O}(10^5)$ iterations, which amounts to $\mcal{O}(10^1)$ seconds of computing time. We have computed stationary solutions for various $\b$ and $\lm$, and some results are presented in \cref{fig_stat}. 

\Cref{fig_stat2} contains information about two key aspects of the stationary polaron state: the electron probability distribution, and the polaron binding energy. Qualitatively speaking, it is in agreement with predictions of the continuum approximation, as per \cref{fig_anal}: as $\lm$ increases, the polaron becomes more localised, and more strongly bound. Moreover, the effect of increasing $\lm$ is more profound given larger values of $\b$. However, further comparison between \cref{fig_anal1,fig_stat2} reveals a noteworthy difference. When $\b = 1$, $\Ph_0$ as a linear function of $\lm$, whereas \cref{fig_stat2} suggests that $\maxpsnsq$, which is approximated by $\Ph_0$, is not linearly dependent on $\lm$. In fact, given any $\b$, $\maxpsnsq$ grows significantly faster with $\lm$ than \cref{fig_anal1} predicts. Despite that, the growth of $\maxpsnsq$ in \cref{fig_stat2} eventually stalls, when $\lm$ becomes sufficiently large. This is a manifestation of a fundamental difference between the continuum and discrete equations, which we discussed in \cref{subsection_anal}: the continuum equations place no limit on how large $\Ph_0$ can be, whereas the discrete system limits $\maxpsnsq$ to 1.

\Cref{fig_stat1} shows a selection of $\psnsq$ solutions. Comparing all the dotted (red) lines, which correspond to $\lm = 1.0$ at various values of $\b$, we see that they are essentially identical. This confirms the belief that when $\lm$ is close to 1, systems with different $\b$-values unify. The figure also shows some stationary solutions to the other half of \cref{dimlesseqns}, namely $u_n$, which is expressed in terms of the stationary $\psnsq$ solution as per \cref{steady_un}. Recall that physically $u_n$ represents the displacement of the $n$\tsups{th} molecule from its equilibrium position. In order that the point-dipole model for lattice units is valid, the lattice distortion must satisfy the condition $\abs{u_{n+1}-u_n} \ll 1$ \cite{Barford2007}. This condition is indeed fulfilled in the stationary polaron state, since according to \cref{fig_stat1} we have $\max \abs{u_{n+1}-u_n} \sim \mcal{O}(10^{-2})$. 

Comparing all the dashed (blue) lines in \cref{fig_stat1}, which correspond to $\lm = 4.4$ at various values of $\b$, enables us to make the following observation. When $\b = 0$, the $u_n$ solution is centred at the location of $\maxpsnsq$, in the sense that its graph is rotationally symmetric about $n=80$. This agrees with our intuition that when $\b = 0$, i.e. when the electron-phonon interaction is spatially symmetric, the electron in the stationary state causes equal lattice distortion to its left and right. As $\b$ increases, the maximum magnitude of $u_n$ remains the same, but the centre of $u_n$ shifts away from the location of $\maxpsnsq$, in response to the decrease in spatial symmetry. When $\b = 1$, the molecule at the location of $\maxpsnsq$ ($n=120$ in this case) is barely displaced, whereas molecules to the right of this point are displaced considerably. Now, the potential energy in the lattice is a sum over terms of the form $(u_{n+1} - u_n)^2$, which is the square of the gradient of the $u_n$ graph at site $n$. In the steady state, this gradient is zero except at a few sites around the location of $\maxpsnsq$, and it is clear that solutions corresponding to larger values of $\b$ have steeper gradients there. We therefore conclude that, in the stationary state, systems with greater spatial asymmetry store more potential energy in the lattice.

In \cref{fig_stat1} we also see a comparison between some $\psnsq$ solutions and their counterpart continuum approximations, $\Ph(x)$. In particular, we look at the thick solid (black) lines and their accompanying thin solid (black) lines. The comparison reveals that, fixing $\lm$, in this case $\lm = 2.6$, $\Ph(x)$ is a more accurate approximant to $\psnsq$ when $\b$ is smaller. As $\b$ approaches 1, it becomes apparent that $\Ph(x)$ under-estimates the height of the $\psnsq$ profile. If $\lm$ is sufficiently large, however, $\Ph(x)$ becomes an over-estimate of the profile height. This all comes down once again to the fact that the continuum equations do not limit the height of the $\Ph(x)$ solution. 

\section{Dynamical polarons in zero temperature} \label{section4}

In this part of the study we explore properties of polarons which propagate along the peptide chain, under an external forcing $\eps(\t)$, and zero temperature ($f_n(\t) = 0$). Physically, $\eps(\t)$ may represent the strength of a time-dependent electric field. We solve \cref{dimlesseqns} as an initial value problem, using the stationary $\ps_n$ and $u_n$ solutions which we computed in \cref{subsection_num} as the initial configuration of the system. We prescribe a suitable $\eps(\t)$, setting $n_0$ to the location where the stationary $\psnsq$ attains its maximum. Then we integrate the system forward in time using the 4\tsups{th}-order Runge-Kutta method. To ensure numerical stability, the integration time-step is set at $\De \t = 0.01$. 

As we forward-integrate the system, we keep track of several scalar quantities associated with the polaron, such as its half-width and binding energy. Most importantly, we keep track of the polaron's position, defined as follows. If $|\ps_n|^2$ attains its maximum at lattice site $\bar{n}$, then polaron position is the vertex location of the parabola extrapolated from three points: $(\bar{n},|\ps_{\bar{n}}|^2),(\bar{n}-1,|\ps_{\bar{n}-1}|^2),(\bar{n}+1,|\ps_{\bar{n}+1}|^2)$. We note that, if the polaron is dynamical, then the stationary solution given by \cref{stat64} is no longer valid, and therefore we cannot take \cref{bindingenergy3} as the expression for the binding energy. Instead, the binding energy $\Eb$ as per \cref{bindingenergy1} will be computed directly from the numerical solutions.

	\subsection{Constant or periodic electric fields} \label{subsection_const}

The most obvious choice of $\eps(\t)$ is a constant,
	\bal
	\eps(\t) = \bareps > 0 \quad \tn{for}~\t \ge 0. \label{const_field}
	\end{align}
Using moderately-localised stationary states (with $\maxpsnsq \approx 0.6$) as initial conditions, we computed polaron trajectories under various values of $\bareps$. Our results show that, given $\b = 1$ and $\lm = 3.0$, a constant forcing of any $\bareps \sim \mcal{O}(10^{-2})$ induces nothing but small oscillations of the polaron around its initial position. An example of trajectory is presented in \cref{fig_EA0_EAA0}. 
	\bg{wrapfigure}{l}{0.57\textwidth}
	\bg{center}
	\includegraphics[width=0.55\textwidth]{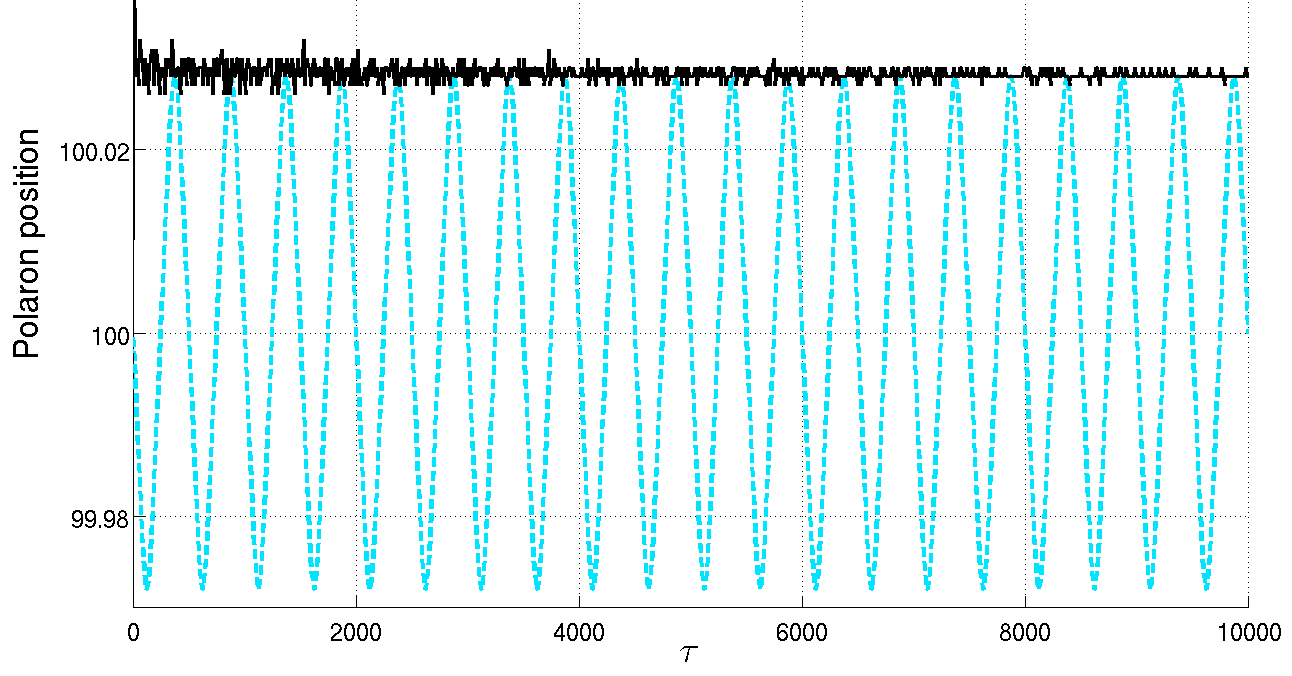}
	\end{center}
	\captionsetup{width=0.5\textwidth}
	\caption{(Colour online.) Some polaron trajectories, given $\b = 1$ and $\lm = 3.0$, under either a constant or a periodic forcing $\eps$. Solid (black) line: $\eps = 0.1$. Dashed (blue) line: $\eps = 0.1 \sin(2\pi \t / T)$, $T =500$. 1000 units of $\t$ equals 1.8 nanoseconds.} \label{fig_EA0_EAA0}
	\end{wrapfigure}
As $\bareps$  is increased beyond 0.1, we find that eventually the forcing does become strong enough to dislodge the electron from its potential well, and propel the polaron along the peptide chain. However, as the polaron propagates, the magnitude of its binding energy decreases rapidly, and the polaron ``delocalises'', i.e. breaks up into unbound components, within several hundred time units. A direct manifestation of the polaron's energy loss and eventual delocalisation is that the $\psnsq$ profile loses height and gains local peaks at lattice sites far away from the global maximum. For the sake of consistency, throughout the remainder of this study we shall say that a polaron has delocalised if its maximum height drops to below 0.1, as it must then be the case that other local peaks have magnitudes comparable to the global maximum. \Cref{fig_EA0_EAA0_2} shows an example of a constant forcing large enough to cause polaron displacement, and it illustrates the resultant rapid delocalisation of the polaron. 

	\bg{figure}[h!]
	\centering
	\bg{subfigure}[h]{0.49\textwidth}
	\captionsetup{width=0.75\textwidth}
	\includegraphics[width=\textwidth]{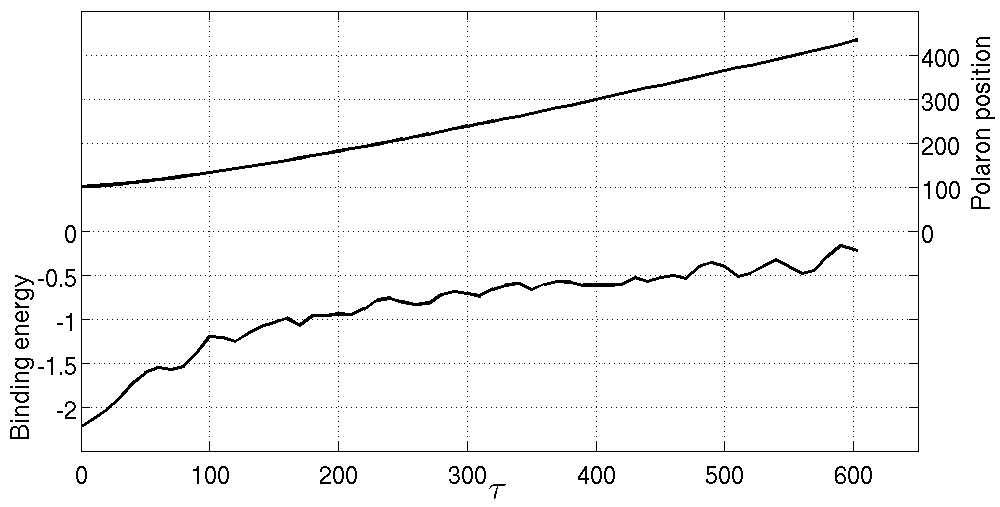}
	\caption{ } \label{fig_EA0_EAA0_scalars}
	\end{subfigure}
	\bg{subfigure}[h]{0.49\textwidth}
	\captionsetup{width=0.75\textwidth}
	\includegraphics[width=\textwidth]{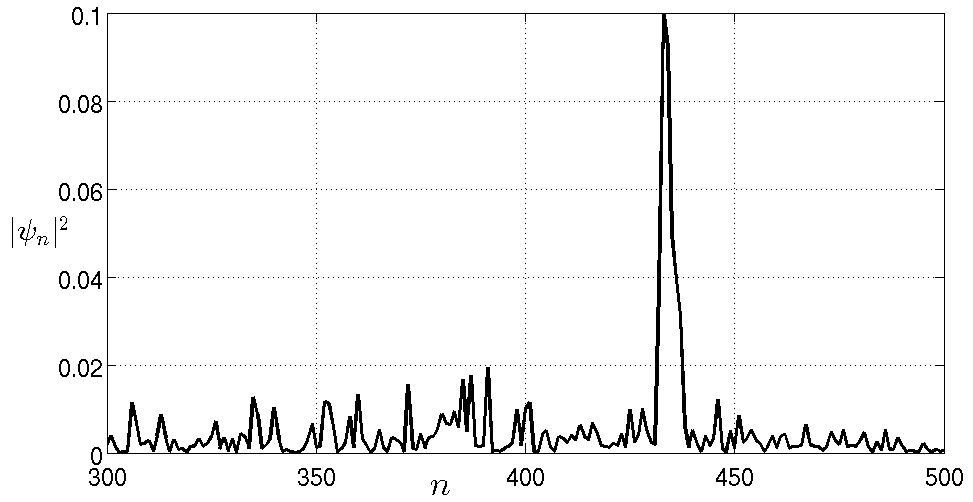}
	\caption{ } \label{fig_EA0_EAA0_a2}
	\end{subfigure}
	\captionsetup{width=0.86\textwidth}
	\caption{Polaron propagation under $\b = 1, \lm = 3.0$ and $\eps = 0.15$. \\ (a) The polaron's position (right axis) and binding energy (left axis) as functions of $\t$. \\ (b) The $\psnsq$ profile of the polaron upon delocalisation. } \label{fig_EA0_EAA0_2}
	\end{figure}
\Cref{fig_EA0_EAA0_scalars} shows the trajectory of a polaron which, within roughly 600 time units, is displaced by just over 300 lattice sites. Its binding energy steadily decreases in magnitude, until the polaron delocalises at $\t \approx 600$. Meanwhile, \cref{fig_EA0_EAA0_a2} shows the electron probability distribution, $\psnsq$, at the time of  delocalisation. This $\psnsq$ profile has evolved from an initial configuration possessing a maximum height of 0.64, and no local peaks apart from the global maximum.

If the polaron's binding energy decreases in magnitude, then the polaron's ability to transport energy is diminished. Beyond the example shown in \cref{fig_EA0_EAA0_2}, all our results are consistent with the hypothesis that, regardless of $\b$ and $\lm$, a constant $\eps$ causes the polaron to undergo either small periodic oscillations, or rapid losses in energy. We would like to find ways to displace the polaron without significant energy loss. Therefore, we must look for forms of $\eps$ other than constants. The next most natural choice of $\eps$ is periodic,
	\bal
	\eps(\t) = A \sin \fr{2\pi \t}{T} \quad \tn{for}~\t \ge 0, \label{per_field}
	\end{align}
where $A$ is the amplitude and $T$ is the period. Physically this may represent an electromagnetic plane wave which is monochromatic, i.e. coherent. Under periodic $\eps(\t)$ with $A$ up to 0.2, regardless of $\b$ and $\lm$ we find that the polaron simply oscillates about its initial position. The polaron's oscillatory motion has a period which coincides with $T$, and an amplitude which is positively correlated with $A$. An example of such trajectories is shown in \cref{fig_EA0_EAA0}. While the polaron remains highly stable over time, its position averaged over its periodic remains constant. Thus, if we want polarons which transport energy from one lattice site to another, we must again look for an alternative form of $\eps(\t)$.

	\subsection{Periodic electric fields with non-zero mean} \label{subsection_harm_const}

Having studied the effects of constant forcing and periodic forcing in \cref{subsection_const}, and discovered that neither serves to displace the polaron with minimal energy loss, in this section we consider forcing of the form
	\bal
	\eps(\t) = \bareps + A \sin \fr{2 \pi \t}{T}. \label{eps_harm_const}
	\end{align}
\Cref{eps_harm_const} represents the combination of the two types of forcing considered previously, with a constant component and a sinusoidal one. One may also think of $\eps(t)$ as a \emph{mean-shifted periodic forcing} (MSPF). In particular, the mean $\bareps$ is chosen to be lower than the constant forcing $\eps$ which is required to displace the polaron, in the manner of \cref{fig_EA0_EAA0_2}. Therefore, $\bareps$ on its own would not give the electron enough energy to escape its potential well. But we hope that the component $A$ can periodically push the electron energy over the threshold, resulting in polaron motion. Another possible advantage of this setup is that $A$ may periodically lower the electron energy, slowing it down and giving the lattice time to ``catch up'', thus making the polaron motion more sustainable than it would be under a constant forcing.

Mathematically, $\eps(\t)$ depends on three independent parameters, $\bareps, A$ and $T$. Before investigating the effect of each of these parameters, we present \cref{fig_EA125_EAA25}, which is a direct comparison with \cref{fig_EA0_EAA0_2}.

	\bg{figure}[h!]
	\centering
	\bg{subfigure}[h]{0.49\textwidth}
	\captionsetup{width=0.75\textwidth}
	\includegraphics[width=\textwidth]{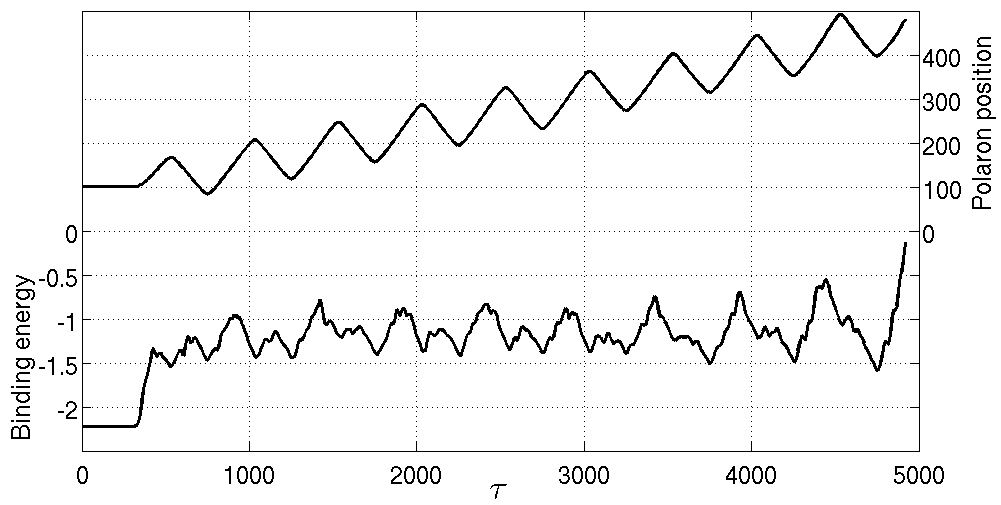}
	\caption{ } \label{fig_EA125_EAA25_1}
	\end{subfigure}
	\bg{subfigure}[h]{0.49\textwidth}
	\captionsetup{width=0.75\textwidth}
	\includegraphics[width=\textwidth]{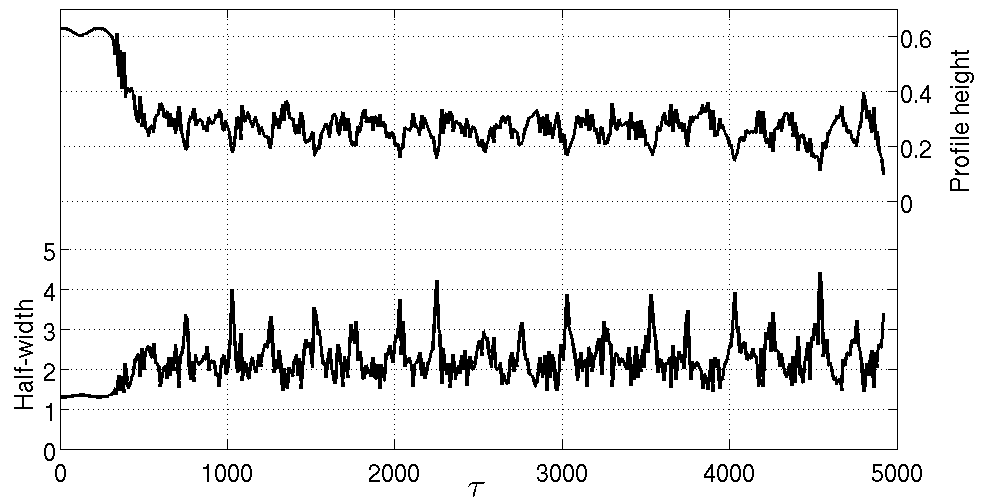}
	\caption{ } \label{fig_EA125_EAA25_2}
	\end{subfigure}
	\captionsetup{width=0.9\textwidth}
	\caption{Polaron propagation under $\b = 1, \lm = 3.0$ and $\eps(\t) = 0.025+0.125\sin(2\pi \t / 500)$. \\ (a) The polaron's position (right axis) and binding energy (left axis) as functions of $\t$. \\ (b) The height (right axis) and half-width (left axis) of the $\psnsq$ profile as functions of $\t$.} \label{fig_EA125_EAA25}
	\end{figure}
We have replaced the constant forcing $\eps = 0.15$, which resulted in \cref{fig_EA0_EAA0_2}, with an MSPF which has the same maximum amplitude as before. The difference is that now this maximum amplitude is reached once every period $T$. \Cref{fig_EA125_EAA25_1} shows that, within roughly 10 periods, the polaron is displaced by nearly 400 lattice sites. Contrary to the uniform manner in which the polaron moves in \cref{fig_EA0_EAA0_scalars}, now the polaron moves towards one end of the peptide chain and then the other, within each period of $\eps(\t)$. The overall displacement of the polaron is due to the fact that each movement to one end of the chain is larger than the subsequent swing back the other way. We note that while the polaron moves slightly further compared to \cref{fig_EA0_EAA0_scalars}, its \emph{lifetime}, i.e. the amount of time elapsed before delocalisation, is much longer. Overall, the polaron in \cref{fig_EA125_EAA25} propagates with a lower \emph{(average) velocity}, $V$, defined by
	\bal
	V = \fr{\tn{average position over final complete period of motion} - \tn{initial position}}{\tn{number of complete periods} \times T }, \label{defn_vel}
	\end{align}  
where the numerator is the \emph{displacement} of the polaron, which we denote by $D$. Our results show that, of the parameters $\bareps,A$ and $T$, the dominant factor which determines the polaron's velocity is the constant component $\bareps$. We will discuss this in more depth in relation to \cref{fig_traj}.

\Cref{fig_EA125_EAA25_1} also shows how the polaron's binding energy, $\Eb$, varies in time. Following an initial drop in magnitude, $\Eb$ mostly oscillates between $-0.75$ and $-1.5$, until another sharp decrease in magnitude leading up to delocalisation at $\t \approx 4900$. There is an important observation to be made here. In \cref{fig_EA0_EAA0_scalars}, we see that when the polaron reaches lattice site $n=300$, $\Eb$ is about $-0.6$. In \cref{fig_EA125_EAA25_1}, the polaron's average position over the 6\tsups{th} period is roughly 300, and the average binding energy over this period is $-1.2$. That is to say, under the MSPF, the polaron is carrying twice as much energy when it reaches $n=300$, compared to when it reaches $n=300$ under the constant forcing. Even though the constant forcing gets the polaron to $n=300$ in less time, we consider the MSPF a better mechanism for polaron transport, because the polaron binding energy is more stable. Indeed, the same can be said when the destination $n$ is anything larger than 30. If the destination is $n < 30$, then the constant forcing takes the polaron to $n$ in such a small amount of time that it causes no more variation in binding energy than the MSPF does. In general, all our results are consistent with the hypothesis that, by splitting a constant forcing into constant and sinusoidal components, we lower the polaron's velocity but increase its stability and lifetime. We say that, compared to the constant forcing, the MSPF is a better \emph{long-distance} tranport mechanism, where speed can be sacrificed for energy efficiency.

In \cref{fig_EA125_EAA25_2} we see another aspect of the polaron's motion, namely, how the height and half-width of the $\psnsq$ profile vary with time. Following an initial decrease, the profile height, $\maxpsnsq$, mostly oscillates between 0.2 and 0.4, until a sudden drop to 0.1, leading to delocalisation. Meanwhile, the half-width mostly oscillates between 1.5 and 4, following an initial growth. The peaks in the half-width, as well as the troughs in $\maxpsnsq$, occur precisely when the polaron turns from moving in one direction to moving in the other. This suggests that when the polaron accelerates, it ``spreads out'', and so the half-width widens and $\maxpsnsq$ drops. We observe this phenomenon in all our results.

	\bg{figure}[h!]
	\centering
	\bg{subfigure}[h]{0.45\textwidth}
	\captionsetup{width=0.9\textwidth}
	\includegraphics[width=\textwidth]{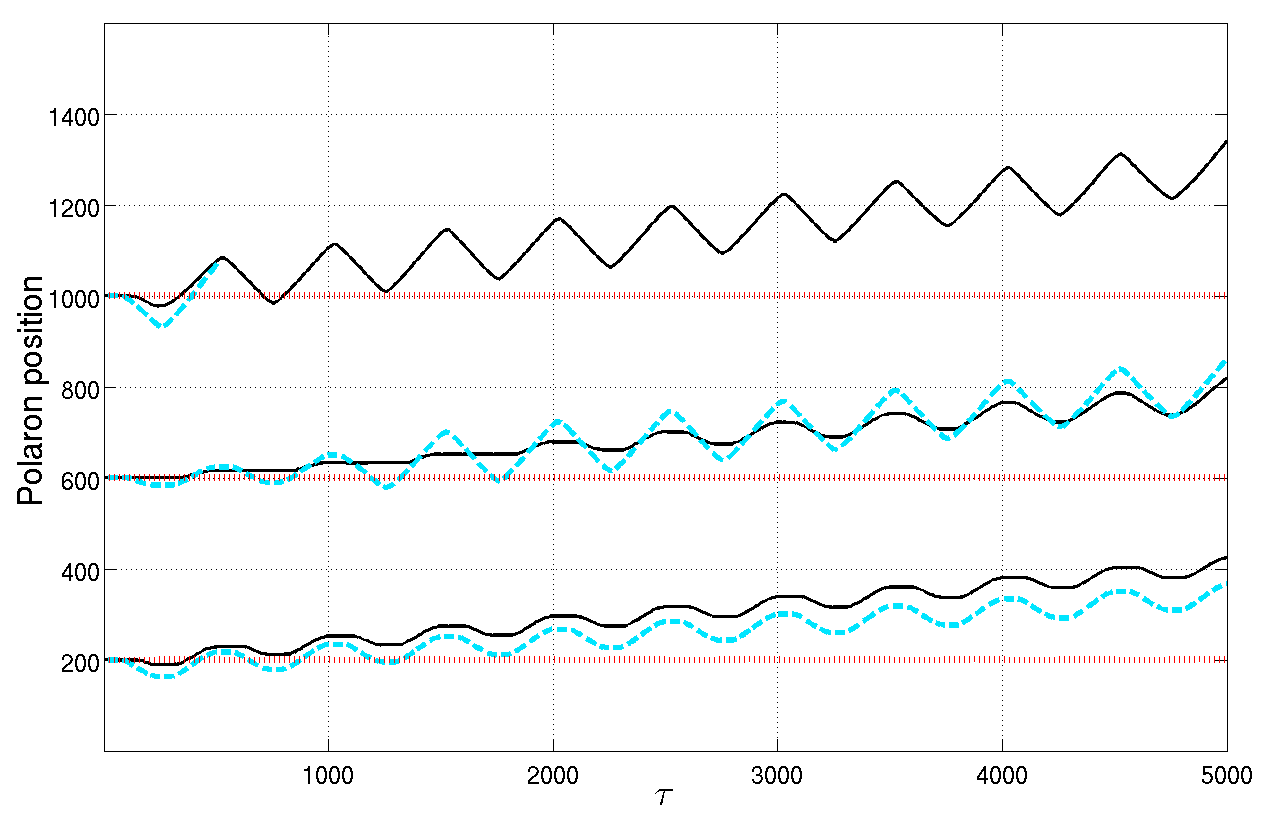}
	\caption{$\bareps = 0.02, T = 500$. } \label{fig_traj1}
	\end{subfigure}
	\bg{subfigure}[h]{0.45\textwidth}
	\captionsetup{width=0.9\textwidth}
	\includegraphics[width=\textwidth]{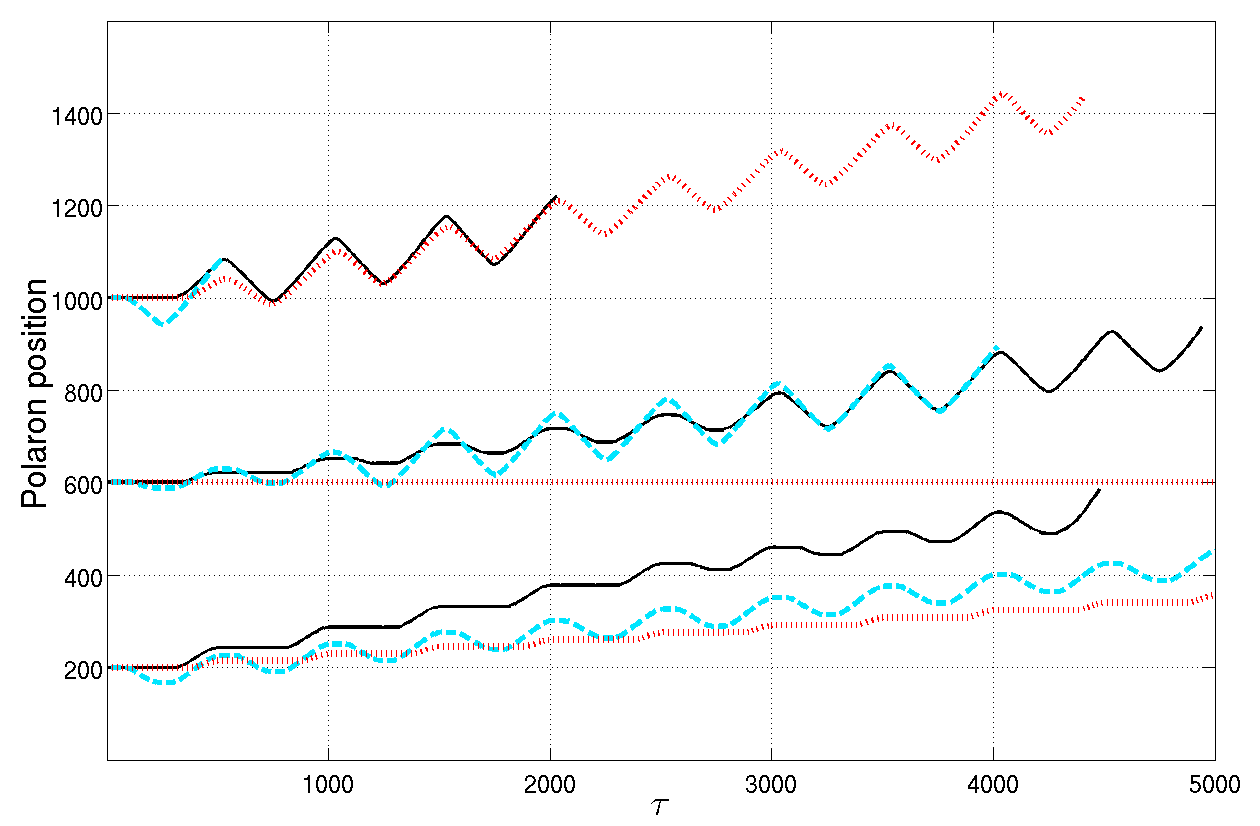}
	\caption{$\bareps = 0.03, T = 500$. } \label{fig_traj2}
	\end{subfigure}
	\bg{subfigure}[h]{0.45\textwidth}
	\captionsetup{width=0.9\textwidth}
	\includegraphics[width=\textwidth]{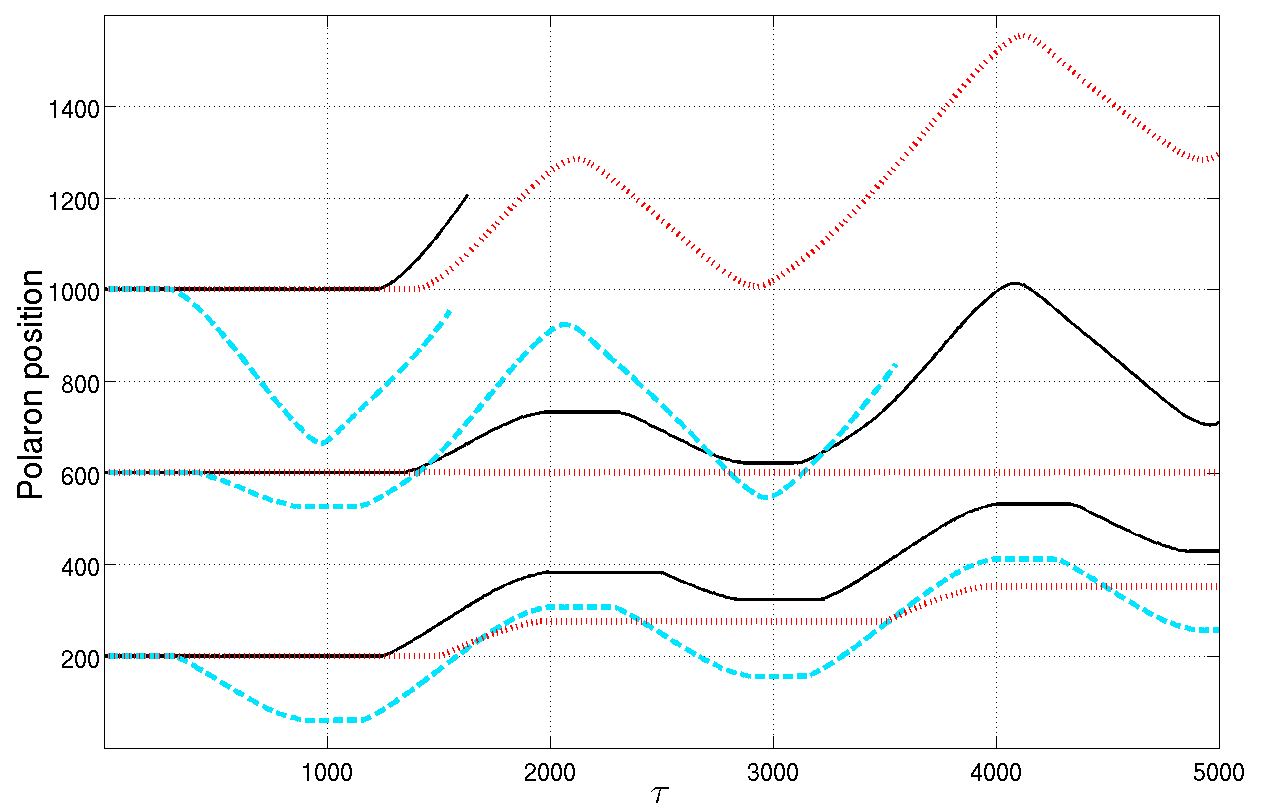}
	\caption{$\bareps = 0.03, T = 2000$. } \label{fig_traj3}
	\end{subfigure}
	\captionsetup{width=0.75\textwidth}
	\caption{(Colour online.) Some polaron trajectories under the MSPF, $\eps(\t) = \bareps + A \sin (2 \pi \t / T)$. Dotted (red) lines: $A = 0.10$; solid (black) lines: $A = 0.15$; dashed (blue) lines: $A = 0.20$. Each figure contains 9 trajectories, whose initial positions have been shifted to avoid overlap. Every trajectory starting from position 200 correspond to a polaron with symmetry parameter $\b = 0$ and effective coupling parameter $\lm =7.6$. Trajectories starting from position 600 correspond to $\b = 0.6$ and $\lm = 4.9$, and those starting from position 1000 correspond to $\b = 1$ and $\lm = 3.0$. $\lm$ has to be varied with $\b$, in order to keep the initial $\psnsq$ profiles unchanged. In this case, all initial conditions have $\maxpsnsq = 0.64$.} \label{fig_traj}
	\end{figure}
Based on our observations, we theorise that a polaron's directed motion may be explained physically as follows. Since the forcing $\eps(\t)$ is the effect of an electric field, it first-and-foremost provides the electron with extra energy. This is evident in the dramatic energy variation during the first period of $\eps(\t)$ (see \cref{fig_EA125_EAA25_1}). Following this, it becomes much easier for the electron to overcome the significantly diminished polaron binding energy, $\Eb$. This is why the onset of polaron motion always follows a drastic drop in magnitude of $\Eb$. Whenever $|\eps(\t)|$ becomes large enough to give the electron sufficient energy to overcome $\Eb$, the electron is dislodged from its potential well and propelled along the lattice. If the electron-lattice coupling is strong enough, then the lattice distortion can keep up with the electron, and so the polaron can remain intact. Whenever $|\eps(\t)|$ drops below the binding threshold, the electron-lattice interaction slows down the electron and causes its probability distribution to spread out. This is why the half-width of $\psnsq$ always peaks at times when the polaron's instantaneous velocity is zero. If $| \eps(\t) |$ remains below the binding threshold for long enough, then the polaron's position can plateau, as is seen in \cref{fig_traj}, particularly in the lowermost solid (black) lines in \cref{fig_traj2,fig_traj3}. If $\eps(\t)$ has a large enough periodic component $A$, then it is possible for $|\eps(\t)|$ to overcome the threshold twice per period: once with $\eps>0$, once with $\eps<0$. If the electron moves towards large $n$ in the $\eps>0$ case, then it will move towards small $n$ in the $\eps<0$ case. This explains the backwards swing exhibited by some polaron trajectories during each period of motion. The fact that $\bareps \neq 0$ ensures that the electron always spends more time moving one way than the other, hence the overall directedness of the polaron trajectories. This point is most clearly demonstrated by the trajectories in \cref{fig_traj3}, where the period $T = 2000$. 

Within each of \cref{fig_traj1,fig_traj2,fig_traj3}, we can compare the trajectories represented by the same line type, but have different starting positions. This reveals the effect of varying the spatial symmetry of electron-lattice interaction. A greater spatial asymmetry, i.e. a larger $\b$, causes the polaron to be more susceptible to displacement. Also within each of \cref{fig_traj1,fig_traj2,fig_traj3}, we can compare trajectories starting from the same position, but are represented by different line types. This reveals the effect of varying the forcing amplitude $A$. The larger $A$ is, the more the polaron oscillates back and forth during each period of motion. We can also compare trajectories in \cref{fig_traj1,fig_traj2} which have identical line types and starting positions. This suggests that the overall velocity of the polaron is determined by $\bareps$, in the sense that the larger $\bareps$ is, the more the polaron moves per unit time. Finally, by comparing trajectories in \cref{fig_traj2,fig_traj3} which have identical line types and starting positions, we hope to see the effect of varying $T$. However, this comparison is not particularly enlightening. We therefore present \cref{fig_paper1}, which not only provides more insight into the effect of $T$, but also helps to quantify our observations, and reinforce our theories, about the effects of $\bareps$ and $A$ . 

	\bg{figure}[h!]
	\centering
	\includegraphics[width=0.96\textwidth]{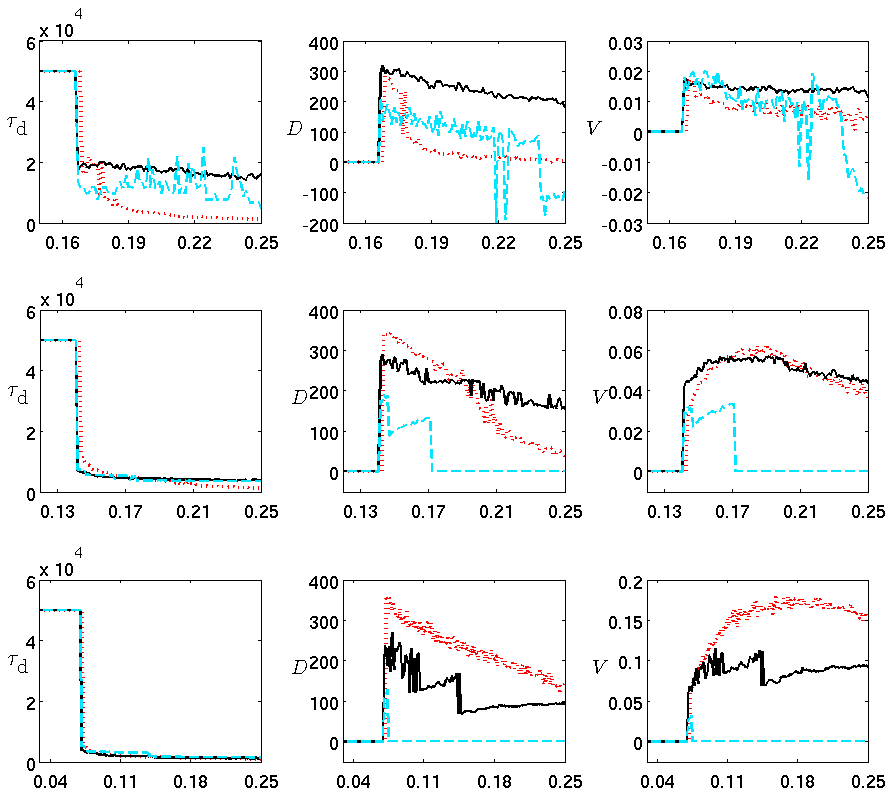}
	\captionsetup{width=0.82\textwidth}
	\caption{(Colour online.) Polaron lifetime $\td$, displacement $D$, and velocity $V$, under the MSPF, $\eps = \bareps + A \sin (2 \pi \t / T)$, with symmetry parameter fixed at $\b = 0.6$. The horizontal axis of each subfigure represents the range of $A$. Top row: $\bareps = 0.005$. Middle row: $\bareps = 0.03$. Bottom row: $\bareps = 0.1$. Dotted (red) lines: $T = 100$. Solid (black) lines: $T = 500$. Dashed (blue) lines: $T = 2000$.} \label{fig_paper1}
	\end{figure}
For several combinations of $\bareps$ and $T$, we examine how the polaron's lifetime, displacement and velocity vary with $A$. The results are displayed together, in \cref{fig_paper1}, so that the effects of $\bareps$ and $T$ can be gauged also. Firtly we consider the lifetime, $\t_\tn{d}$. We computed all our numerical solutions up to $\t = 50000$, which is several times larger than the typical lifetime of a polaron that moves under the MSPF. If the polaron is not displaced by the MSPF, then it is effectively permanent, in the sense that its energy oscillates instead of dissipating over time, and it would have a lifetime far exceeding 50000. Thus, in \cref{fig_paper1}, the lifetime of a permanent (undisplaced) polaron is represented as $\td = 50000$. For each combination of $\bareps$ and $T$, there exists some \emph{critical amplitude}, $A = \Ac$, below which the polaron is undisplaced by the MSPF. At $A = \Ac$, the combined magnitude of the forcing, $\epscomb := \bareps + A$, becomes large enough to displace the polaron, and $\td$ drops sharply. This drop can sometimes result in a lifetime of only several thousand time units - see for instance the bottom-left subfigure in \cref{fig_paper1}, corresponding to $\bareps = 0.1$. When $\bareps$ is smaller, say $\bareps = 0.005$ (top-left subfigure), the drop in lifetime is less dramatic. As $A$ increases beyond $\Ac$, the polaron's lifetime drops further, if only slightly. 

Next, we look at the polaron's displacement, $D$. When $A$ is small, the polaron does not move barring small oscillations, the types of which we saw in \cref{fig_EA0_EAA0}. As $A$ reaches critical value $\Ac$, the polaron turns from being quasi-stationary to moving by several hundred lattice sites during its lifetime. Evidently, the value of $\Ac$ is independent of $T$. Note that we only consider the displacement of polarons whose lifetimes are at least $2T$, and we set the displacement of polarons with shorter lifetimes to zero - see for instance the dashed (blue) lines in the centre and bottom-middle subfigures.

Whilst the value of $\Ac$ does not depend on $T$, the amount of displacement caused by $\Ac$ does. However, it is unclear from our results what their correlation is. As $A$ increases beyond $\Ac$, the qualitative behaviour of $D$ is that it decreases. This is due to the fact that increasing $A$ causes the polaron to delocalise more quickly, and therefore the polaron has less time to move. To understand how $A$ affects the amount of polaron displacement \emph{per unit time}, we examine the polaron's (average) velocity, $V$, as per definition (\ref{defn_vel}). When $A$ is small, $V$ is zero. As $A$ reaches critical value $\Ac$, the velocity becomes typically $\mcal{O}(10^{-2})$. Exactly what value this \emph{critical velocity} $\Vc$ takes depends on $\bareps$ - the larger $\bareps$ is, the larger $\Vc$ is. As $A$ increases beyond $\Ac$, sometimes $V$ simply decays away - see for instance the top-right subfigure, where $\bareps = 0.005$. Sometimes, however, $V$ grows before its decay - see for instance the middle-right and bottom-right subfigures, where $\bareps = 0.03$ and 0.1 respectively. Such behaviour is possible when the polaron lifetime decays with $A$ more quickly than the displacement does. When this happens, there may exist some \emph{optimal amplitude}, $A = \Am$, at which the polaron attains maximum velocity, $\Vm$. $\Am$ may coincide with $\Ac$ - see for instance the top-right subfigure. Meanwhile, the middle-right and bottom-right illustrate clearly that, for different values of $T$, the critical $\Ac$ remains the same, whereas the optimal $\Am$ changes. Qualitatively speaking, the larger $T$ is, the smaller $\Am$ is. 

Whilst the value of $\Ac$ does not depend on $T$, it does depend on $\bareps$ - we see this by comparing any row of subfigures in \cref{fig_paper1} to any other row. But how do $\Ac$ and $\bareps$ correlate? Our results show that, as $\bareps$ grows, $\Ac$ drops, but crucially the combined magnitude $\epscomb = \bareps + \Ac$ remains roughly constant. Specifically, in the top row we see $\bareps = 0.005$ and $\Ac = 0.167$, in the middle row we have $\bareps = 0.030$ and $\Ac = 0.142$, and the bottom row shows $\bareps = 0.100$ and $\Ac = 0.072$, each case giving $\epscomb = 0.172$ when $A$ reaches critical. Recall that, when using a straightforward constant forcing $\eps = \bareps$, there is also a threshold value for $\bareps$, below which the polaron simply exhibits small oscillations, and above which the polaron moves at high speed but delocalises very quickly. It is noteworthy that this threshold is $\eps = 0.154$ (given $\b = 0.6$), which is significantly lower than the critical combined amplitude of 0.172. In other words, $\eps = 0.154$ causes polaron displacement, $\eps = 0.153$ does not; and if one wishes to add on a periodic component $A\sin(2 \pi \t / T)$ in order to move the polaron, one needs $A \ge 0.019$, making $\bareps + A$ far exceed what $\bareps$ is required on its own to move the polaron. This phenomenon is observed across all values of $\b$.

In practice, then, what would make a good combination of forcing parameters, which propel the polaron with decent speed but does not cause large energy dissipation too quickly? First of all, a large $\bareps$ results in a large velocity but an energetically unstable polaron which delocalises rapidly - so rapidly that it may move the polaron less far in its lifetime than a small $\bareps$ does. The middle column of \cref{fig_paper1} precisely illustrates this point. Meanwhile, a small $\bareps$ results in long-living polarons which can move very far, because of how stable they are, but they would take more time to reach the same destination, compared to polarons under a large $\bareps$. On balance, a moderate value of $\bareps$ such as 0.03 is preferable. Secondly, once a $\bareps$ is chosen, it remains to choose $A$ and $T$, and it is obvious that the ideal choice of $A$ is the optimal amplitude, $A=\Am$. Meanwhile, if $T$ is small, such as $T=100$, $\Am$ would be large. On the other hand,  if $T$ is large, such as $T = 2000$, the value of the maximum velocity would be small. We observe both of these extremes very clearly in the middle-right subfigure of \cref{fig_paper1}. Once again, these observations are not specific to $\b = 0.6$, but universal for all values of $\b$. Overall, we believe that the best MSPF parameters which we have tested are such combinations where $\bareps \approx 0.030$, $T \approx 500$, and $A \approx \Am$ which, given $\b = 0.6$, is $\Am = 0.157$. We will discuss the relationship between $\Am$ and $\b$ in \cref{subsection_beta}.

We note one anomaly which we observe in \cref{fig_paper1} but did not expect. When $\bareps = 0.005$ (top row), if $T = 2000$ and $A$ is large enough, then the displacement (and therefore velocity) can take large negative values, meaning the polaron moves in the opposite direction to what we expected, and with large speeds. Whilst we are uncertain as to what causes this counter displacement, it is certainly another reason to reject small $\bareps$ and large $T$ when choosing forcing parameters.

	\subsection{The relevance of $\b$} \label{subsection_beta}

To produce \cref{fig_paper1}, we fixed $\b = 0.6$. How would the figure have looked if $\b$ had been different? Our results show that qualitatively it would exhibit the same behaviour, characterised by critical amplitudes $\Ac$, and optimal amplitudes $\Am$. Quantitatively, the values of $\Ac$ and $\Am$ would change. It is therefore natural to investigate how they change with $\b$. After all, our generalisation to the Davydov-Scott model is manifest in the extra parameter $\b$. Firstly we establish the following preliminary result. 

Recall that the stationary polaron, upon which we impose the MSPF, is characterised by two quantities: its probability distribution, specifically its maximum localisation probability, $\maxpsnsq$, and its binding energy. These are in turn determined by the symmetry parameter $\b$ and effective coupling parameter $\lm$. As $\b$ varies, so does the value of $\lm$ required to keep $\maxpsnsq$ constant. This correlation is shown in \cref{fig_beta_init}. It is clear that $\lm(\b)$ is a decreasing function. We have made sure that whenever we altered $\b$ we also took $\lm = \lm(\b)$, so that all of our moving polarons begin as stationary states which share the same probability distribution. An alternative would have been to take whatever $\lm$ is required to keep the binding energy constant. Our results show that if we had decided to keep the binding energy constant at, say, $-2.5$, then $\maxpsnsq$ would have been 0.53 at $\b=0$, or  0.81 at $\b=1$. It is a central feature of our model that two stationary polarons with the same maximum localisation probability need not have the same binding energy, and vice versa. 

	\bg{figure}[h!]
	\centering
	\bg{minipage}{0.49\textwidth}
	\centering
	\includegraphics[width=0.86\textwidth]{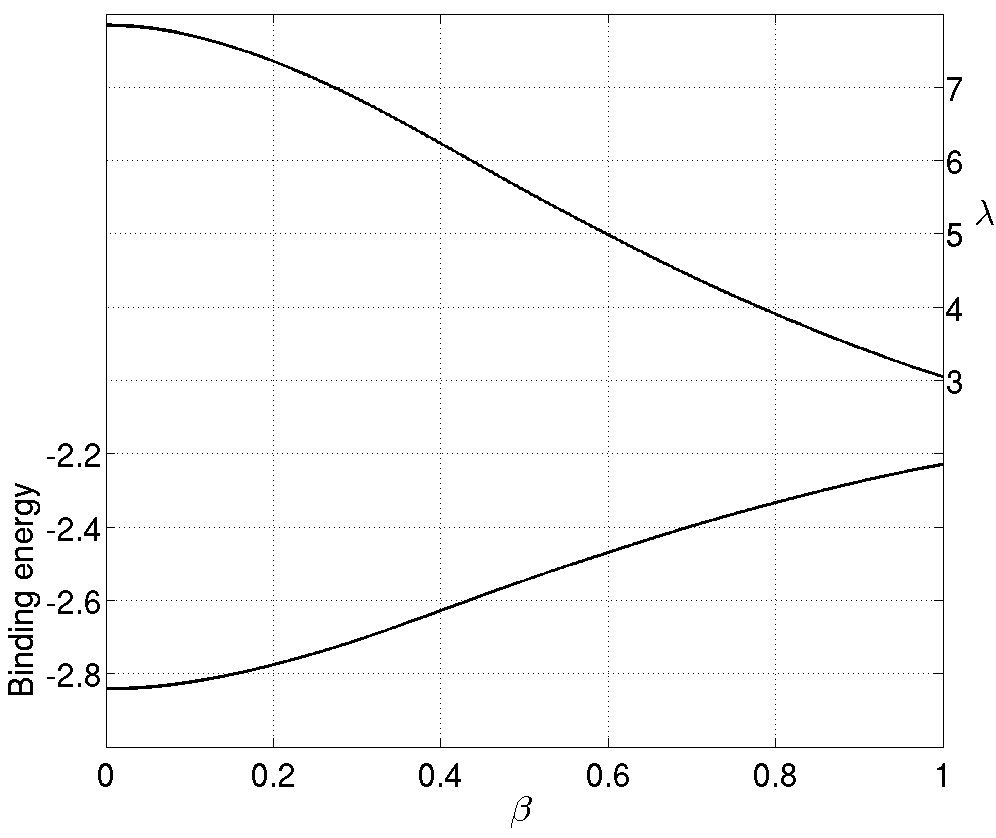}
	\captionsetup{width=0.75\textwidth}
	\caption{For $\b$ in $[0,1]$: right axis: $\lm(\b)$, the value of $\lm$ required to keep $\maxpsnsq = 0.64$; left axis: binding energy of stationary polaron resulting from $\b$ and $\lm(\b)$. For example, when $\b = 0.6$ and $\lm = \lm(0.6) = 5.0$, the binding energy is $-2.47$.} \label{fig_beta_init}
	\end{minipage}%
	\bg{minipage}{0.49\textwidth}
	\centering
	\includegraphics[width=0.96\textwidth]{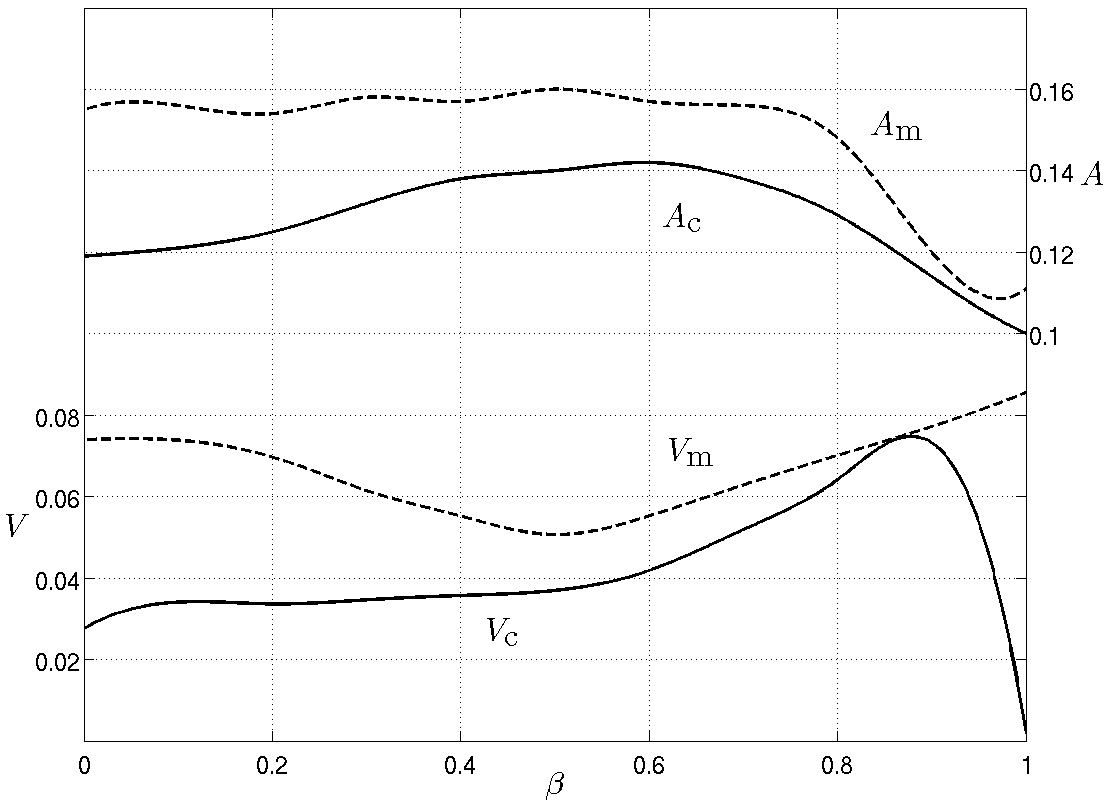}
	\captionsetup{width=0.75\textwidth}
	\caption{Critical amplitude $\Ac$ (right axis, solid line), optimal amplitude $\Am$ (right axis, dashed line), critical velocity $\Vc$ (left axis, solid line), and optimal velocity $\Vm$ (left axis, dashed line), as functions of $\b$. Parameters: $\bareps = 0.03, T = 500$.} \label{fig_beta}
	\end{minipage}
	\end{figure}
Having established the relation $\lm(\b)$, we can study how $\Ac$ and $\Am$ depend on $\b$. We do so by fixing $\bareps$ and $T$, and working out what $\Ac$ and $\Am$ are for various $\{ \b, \lm(\b) \}$. For instance, in \cref{fig_paper1} we saw that if $\b = 0.6$, $\lm = \lm(0.6) = 5.0$, $\bareps = 0.03$ and $T = 500$, then $\Ac = 0.142$ and $\Am = 0.157$. What if we fix $\bareps$ and $T$, and vary $\b$? The result is displayed in \cref{fig_beta}. When $\b = 0$, we have $\Ac = 0.12$, and when $\b = 1$ we have $\Ac = 0.1$. In fact, $\Ac$ is minimal when $\b = 1$, which suggests that a system with antisymmetric electron-phonon interaction is most conducive to polaron displacement by MSPF. One might expect that a system with symmetric interaction would be least conducive to polaron displacement, and therefore $\Ac$ should be maximal when $\b = 0$. This is not the case. We observe that $\Ac$ is maximal when $\b = 0.6$, which models a system with moderately asymmetric electron-phonon interaction. Meanwhile, polaron velocity produced by critical forcing, $\Vc$, is maximal when $\b=0.87$. 

The optimal amplitude, $\Am$, varies little when $\b$ is less than 0.7, but decays sharply when $\b$ increases beyond 0.7, to such an extent that it almost equals the critical amplitude $\Ac$. The optimal velocity, $\Vm$, is typically of the same order of magnitude as $\Vc$. When $\b = 0.87$, $\Vm$ and $\Vc$ almost coincide.

Earlier, based on \cref{fig_paper1}, we asserted that the period $T$ of the MSPF has little effect on the value of $\Ac$. Indeed, our results show that, if we had produced \cref{fig_beta} with $T$ fixed at either 100 or 2000, the $\Ac$ curve would have been virtually unaffected. We also conjectured that the critical amplitude $\Ac$ is negatively and linearly correlated with $\bareps$, so that the combined magnitude $\epscomb = \bareps+\Ac$ remains constant as $\bareps$ varies. This is supported by our results. Indeed, \cref{fig_beta} is produced with $\bareps$ fixed at 0.03; but if we had produced \cref{fig_beta} with $\bareps$ fixed at either 0.005 or 0.1, the $\Ac$ curve would simply have been shifted along the vertical axis, by an amount equal to the difference between the new $\bareps$ and 0.03.

	\section{Dynamical polarons in non-zero temperature} \label{section5}

We study the effect of random fluctuations which result from non-zero temperatures in the environment surrounding the lattice. The randomised forcing on the lattice is represented by the normally-distributed $f_n(\t)$, which by definition (\ref{dimlessparams}) must have, for $\t \ge 0$, the following first and second moments.
	\bsubs \label{fnmoments}
	\bal
	\brk{f_n(\t)} &= 0, \label{zero_mean} \\
	\brk{f_m (\t)f_n (\t + \De \t)} &= \fr{2 \gamma \th \de_{mn} }{\De \t}, \label{fmfn_corr}
	\end{align} 
	\esubs 
where $\th$ is the dimensionless temperature,
	\bal
	\th = \fr{k_B \Theta}{MR^2\O^2}, \label{defn_theta}
	\end{align}
and $\Theta$ is the temperature. Compared to the dimensional $F_n(t)$ that we introduced in \cref{section2}, the $\t$ which now appears in $f_n(\t)$ is a discrete index. We have followed the standard procedure of replacing $\de(t - t')$ by $1 / \De \t$, up to non-dimensionalisation constants. Beginning with a stationary polaron, which we computed numerically in \cref{subsection_num}, we integrate the system of \cref{dimlesseqns} forward in time from $\t = 0$. Using a random number generation algorithm, we generate a new vector $f_n$ before each integration step. If $\th$ is large, we find that it can cause large distortions in the lattice and rapid delocalisation of the polaron, due to excessive energy input to the system. For appropriate values of $\th$, we see that the polaron's binding energy undergoes small fluctuations, but on the average it tends to shift towards zero. After some time, the binding energy stabilises. For example, given $\b = 0.6$ and $\lm = 5.0$, the stationary polaron has binding energy $-2.47$. Integrating from $\t = 0$ with $\th = 0.0003$, we find that after $\t \sim \mcal{O}(10^4)$ the binding energy settles, on average, around $-2.15$. The period of time required for a polaron to reach such a thermal equilibrium is the \emph{thermalisation phase} of the polaron dynamics. During this phase, the forcing $\eps(\t)$ on the electron is kept at zero, but the electron nevertheless undergoes small fluctuations around its initial position, due to its coupling to the thermalised lattice. Our results show that, irrespective of $\b$, we are unable to raise $\th > 0.001$, because such a large $\th$ induces excessive lattice distortions which cause the polaron to delocalise before reaching thermal equilibrium. 

In each simulation, we integrate the system, with $\eps(\t) = 0$, until the polaron reaches thermal equilibrium. Then we reset $\t = 0$, and ``turn on'' the forcing $\eps(\t)$ for $\t \ge 0$. We examine how the polaron subsequently moves, under combinations of $\eps(\t)$ and $f_n(\t)$. Since the thermalisation phase raises the polaron energy, we expect that a thermalised polaron would be easier to displace, in the sense that it would require a smaller $\eps(\t)$ to displace it, compared to the zero temperature case. Indeed, our results confirm this. To obtain our results in this section, every dynamical simulation, with a set of chosen parameters $\{ \b, \lm(\b), \bareps, A, T, \th \}$, is run 100 times, and averages of quantities such as polaron lifetime and displacement are then taken. 

	\bg{figure}[h!]
	\centering
	\includegraphics[width=0.85\textwidth]{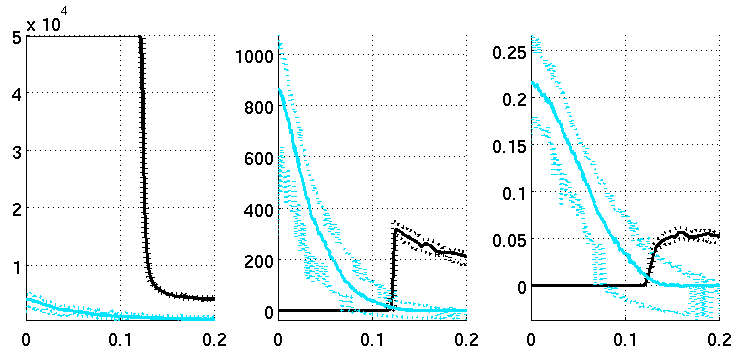}
	\captionsetup{width=0.8\textwidth}
	\caption{(Colour online.) From left to right: polaron lifetime $\td$, displacement $D$, and velocity $V$, under the MSPF, $\eps = \bareps + A \sin (2 \pi \t / T)$, and the thermal forcing, $f_n(\t)$ with temperature $\th$. The horizontal axis is $A$. The symmetry parameter fixed at $\b = 0.6$. $\bareps = 0.03$ and $T = 500$ are fixed. Black lines: $\th = 0.0001$. Grey (blue) lines: $\th = 0.0005$. Each simulation of polaron dynamics is run 100 times, and the average result is shown in solid lines, while the maximum or minimum results are shown in dotted lines. } \label{fig_thermal_beta06}
	\end{figure}
	\bg{figure}[h!]
	\centering
	\bg{minipage}{0.49\textwidth}
	\centering
	\includegraphics[width=0.96\textwidth]{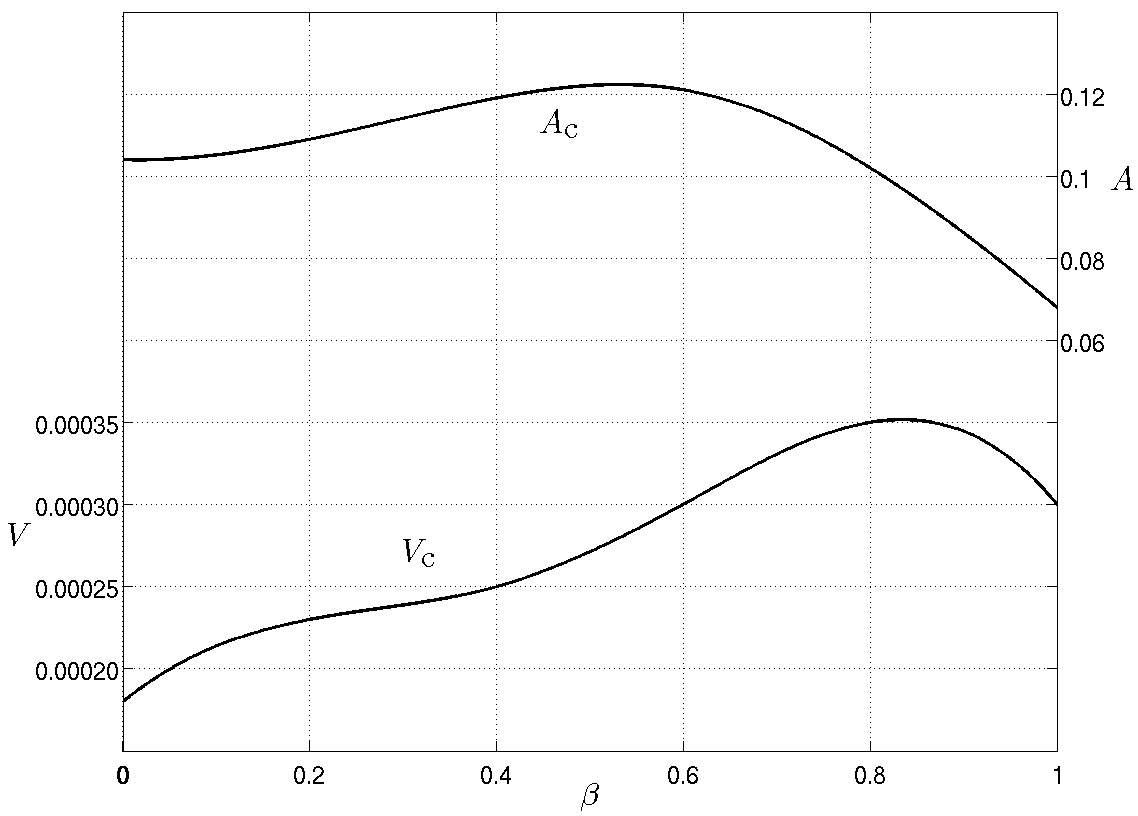}
	\captionsetup{width=0.75\textwidth}
	\caption{Critical amplitude $\Ac$ (right axis) and critical velocity $\Vc$ (left axis),  as functions of $\b$. Parameters: $\bareps = 0.03, T = 500, \th = 0.0001$.} \label{fig_thermal_beta_THETA1e_4}
	\end{minipage}%
	\bg{minipage}{0.49\textwidth}
	\centering
	\includegraphics[width=0.71\textwidth]{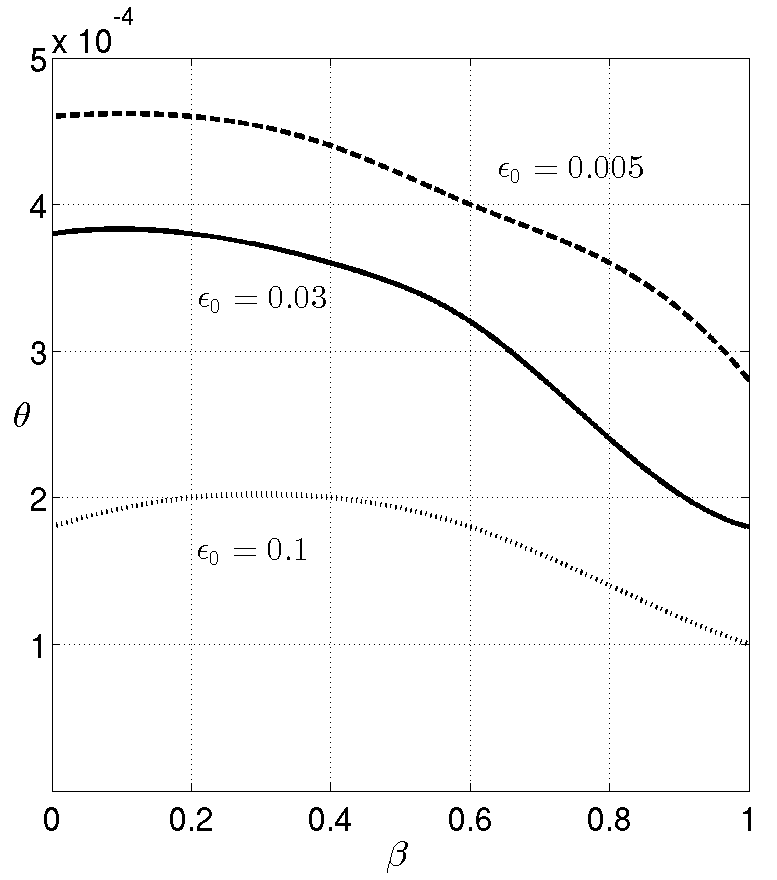}
	\captionsetup{width=0.75\textwidth}
	\caption{Critical temperature $\thc$ as a function of $\b$ and $\bareps$, $A = 0$.} \label{fig_thermal_detail5}
	\end{minipage}
	\end{figure}
\Cref{fig_thermal_beta06} is to be compared directly with \cref{fig_paper1}, which contained results for $\b = 0.6$ and $\th = 0$. Specifically, \cref{fig_thermal_beta06} is to be compared with the solid (black) lines in the middle row of subfigures in \cref{fig_paper1}, for which two of the parameters in $\eps = \bareps + A \sin (2 \pi \t / T)$ were fixed: $\bareps = 0.03$, and $T = 500$. We saw that, given said parameter values, the critical amplitude was $\Ac = 0.142$. When we have a non-zero $\th$ in the system, we define $\Ac$ to be the smallest $A$ for which the average polaron displacement (over 100 simulations) exceeds 10 lattice sites. According to this definition, when $\b = 0.6, \bareps=0.03, T = 500$ and $\th = 0.0001$, we see in \cref{fig_thermal_beta06} that $\Ac = 0.121$, which is significantly lower than the case of $\th = 0$. Fixing said values of $\bareps, T$ and $\th$, we find that the value of $\Ac$ depends on $\b$ in a manner shown in \cref{fig_thermal_beta_THETA1e_4}. That is, $\Ac$ is minimal when $\b = 1$, suggesting that an antisymmetric electron-phonon interaction makes it easiest to displace the polaron. Meanwhile, $\Ac$ is maximal when $\b \approx 0.5$, suggesting that, counter-intuitively, what makes displacing the polaron most difficult is not a symmetric electron-phonon interaction, but a moderately asymmetric one. Indeed, this $\Ac(\b)$ function is very similar to the one in \cref{fig_beta}, where we also had $\bareps = 0.03, T = 500$ fixed, but $\th = 0$. Now with $\th = 0.0001$, the $\Ac(\b)$ curve in \cref{fig_thermal_beta_THETA1e_4} is significantly lower. This means that, regardless of $\b$, a non-zero temperature makes it easier to displace the polaron by the forcing $\eps = \bareps + A \sin (2 \pi \t / T)$, in the sense that a smaller combined magnitude $\bareps + A$ is required. It is also noteworthy that, under a non-zero temperature, the onset of polaron motion is more gradual, in the sense that a critical amplitude results in a very small velocity, $\Vc$. Indeed, comparing $\Vc$ in \cref{fig_beta} with $\Vc$ in \cref{fig_thermal_beta_THETA1e_4}, we see that the latter is 2 orders of magnitude smaller.

More can be said about \cref{fig_thermal_beta06}. When we raise the temperature to $\th = 0.0005$, we find that the polaron is displaced (on average) by hundreds of sites even if $A = 0$.  This suggests that, given $\b = 0.6$ and $\bareps = 0.03$, there exists some \emph{critical temperature} $\th = \thc$ between 0.0001 and 0.0005, for which just the combination of $\eps(\t) = \bareps$ and $f_n(\t)$ is sufficient to displace the polaron, and no periodic component in $\eps(\t)$ is needed. $\thc$ is critical in the sense that, if $\th$ is any lower than $\thc$, then the combination of $\eps(\t) = \bareps$ and $f_n(\t)$ does not energise the polaron enough to move it, and a non-zero $A$ is required. Indeed our results show that, given $\b = 0.6$ and $\bareps = 0.03$, the critical temperature is $\thc = 0.00032$. Furthermore, we have investigated how $\thc$ changes as we vary $\b$ and $\bareps$, and the results are shown in \cref{fig_thermal_detail5}. We observe the qualitative trend that, the larger $\bareps$ is, the less thermal energy is required to make up for the extra energy that the polaron needs in order to move. We also observe that, in general, the larger $\b$ is, the less thermal energy is required to displace the polaron. This fits in nicely with our understanding that, when $\b$ is close to 1, we have an electron-phonon interaction which is biased towards one end of the lattice, making the electron more susceptible to displacement. 
	\bg{wrapfigure}{l}{0.57\textwidth}
	\bg{center}
	\includegraphics[width=0.55\textwidth]{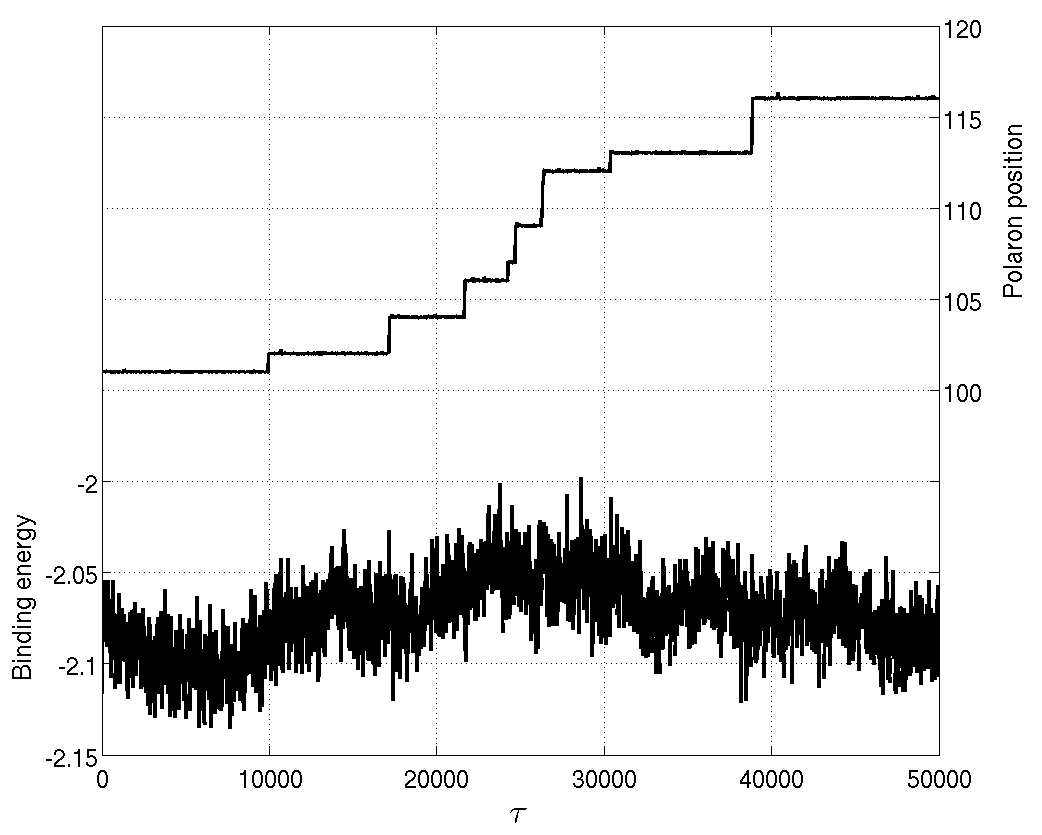}
	\end{center}
	\captionsetup{width=0.52\textwidth}
	\caption{A polaron trajectory (right axis), and the corresponding time-evolution of the polaron's binding energy (left axis), given $\b = 0.6, \bareps = 0.03, A=0$, and $\th = \thc = 0.00032$.} \label{fig_thermal_traj2}
	\end{wrapfigure}
In \cref{fig_thermal_traj2} we present a typical polaron trajectory when $\th = \thc$. Under this critical temperature, some simulations would produce no polaron displacement at all, but most trajectories would be similar to that in \cref{fig_thermal_traj2}, clearly showing a directed movement. We propose to explain the shape of these trajectories as follows. First of all, the combined magnitude of the MSPF would be much lower than what is required to move the polaron under zero temperature. In \cref{fig_thermal_traj2} for example, we have $\bareps + A = 0.03$, whereas the critical value under zero temperature, as we discovered in \cref{subsection_harm_const}, is $\bareps + A = 0.172$. Even when there is a non-zero temperature, the polaron still spends the majority of its lifetime oscillating around its localisation site by small amounts. However, occasionally the random forces on the lattice sites in the vicinity of the electron causes a large distortion, such that the effective potential barrier for the electron is significantly lowered, and the electron can escape the well. Once it does that, it is propelled towards one end of the lattice by $\bareps$. But before the electron has time to move far, the random forces may have further distorted the local lattice sites in such a way that a high potential barrier is restored. This then traps the electron again, giving the polaron time to recover its integrity, before the next random time at which the electron jumps out of its potential well. This explains why a trajectory under the critical temperature appears jagged, showing the polaron ``hopping'' one or two sites at a time, in stark contrast with the smooth and regular polaron motion exhibited in \cref{fig_traj}. 

	\section{Discussions and conclusions} \label{conclusions}

In this study we have presented a new mathematical model describing polaron dynamics in linear peptide chains. The model is dependent on a symmetry parameter, $\b$, which measures the extent to which the interaction between the polaron's electron and phonon components is spatially symmetric. We have shown that when $\b$ takes its extreme values, 0 and 1, the model reduces to existing ones for which it was assumed that the electron-phonon interaction was, respectively, symmetric and antisymmetric. We have justified the physical neccessity of including $\b$ in the model, in that one should not simply assume the electron to be coupled equally strongly to lattice points on either side, or to be coupled only to the lattice point on one side. Instead, the spatial symmetry should be determined by the adjustable parameter $\b$. 

Apart from $\b$, we have also identified two composite parameters which are most vital to the intrinsic properties of the polaron. Firstly there is the adiabaticity parameter, $\rh$, measuring the characteristic time scale separation between the electron and phonon, which we justifiably fixed throughout the study. Then there is the effective coupling parameter, $\lm$, measuring the strength of the electron-phonon interaction. The combination of $\b$ and $\lm$ determines the two aspects of the stationary polaron: its maximum localisation probability, and its binding energy. We have computed both of these quantities as functions of $\b$ and $\lm$. Moreover, in the infinite lattice limit, we have obtained stationary polaron solutions by analytically integrating the system, and the results are in good agreement with our numerical solutions on a finite lattice. 

Our main results relate to using an external forcing to displace the polaron, in a manner which causes minimal energy loss and which, crucially, is directed. Such polaron dynamics could be achieved only if the electron is dislodged from its self-trapping potential well, and the local lattice distortions propagate coherently with the electron, and some mechanism exists which ensures the electron always moves towards one end of the lattice. If the second condition is not met, then over time the electron probability density function would become broader, leading to delocalisation of the polaron. We have found that a constant external force, $\bareps$, on the electron is insufficient for displacing the polaron, unless $\bareps$ is larger than some threshold value, but then the forcing causes rapid energy loss and delocalisation. We have also found that a sinusoidal force, $A \sin (2 \pi \t / T)$, on the electron is never sufficient for displacing the polaron, throughout the range of $A$ that we tested. We then combined the constant and sinusoidal forces, resulting in the mean-shifted periodic forcing (MSPF), $\eps = \bareps + A \sin (2\pi \t /T)$. We have discovered that, for each $\bareps$ which is insufficient on its own to displace the polaron, there is some critical value $\Ac$, such that the polaron is displaced if and only if $A \ge \Ac$. There is also an optimal value $\Am$, such that the polaron attains maximum velocity at $A = \Am$. The value of $\Ac$ is irrespective of the period $T$, whilst $\Am$ is negatively correlated with $T$. As $\bareps$ is decreased, $\Ac$ increases, in such a way that the combined magnitude $\bareps + \Ac$ remains constant. This suggests that there is a certain amount of extra energy that the electron needs in order to overcome the polaron binding, and how much of it comes from the constant or sinusoidal part is inconsequential, as long as the two parts combine to a large enough overall amplitude. Nevertheless, the split between $\bareps$ and $A$ does determine the manner in which the polaron propagates, specifically its velocity and stability. The velocity is predominantly determined by $\bareps$, and positively correlated with it; but the stability of the polaron is negatively correlated with $\bareps$. By comparing three sets of $\{\bareps, A\}$ with the same combined $\bareps+A$, namely $\{0.005, 0.167\}, \{ 0.03, 0.142\}, \{0.1, 0.072\}$ (while keeping all other parameters fixed), we found that $\{0.03,0.142\}$ produces optimal balance between polaron velocity and stability. 

We have examined how the aforementioned phenomena depends upon $\b$. To do so, we needed a way of isolating the effect of varying $\b$. This posed a difficulty, because if we were to fix $\lm$ and vary $\b$ then both the maximum localisation probability and binding energy of the stationary polaron would change. We would then be comparing dynamical behaviours of dissimilar polarons. We therefore decided to vary $\lm$ with $\b$, in a way that allowed us to generate a set of stationary polarons, one for each combination of $\{ \b, \lm(\b) \}$, such that they all had the same maximum localisation probability. Then we launched these polarons using the same external forcing and compared the results. We have found that $\b = 1$, representing a spatially antisymmetric electron-phonon interaction, produces a polaron which is easiest to move, in the sense that the least amount of forcing is required. We have also found that the symmetric model, $\b = 0$, does not make the polaron most difficult to move ($\b \approx 0.6$ does that). This hints at the existence of some intrinsic mechanism in the $\b = 0$ model which pushes the electron towards one end of the lattice, despite it being coupled to the other end equally strongly. 

We have also studied the MSPF under non-zero temperatures, $\th > 0$. The manifestation of thermal effects is random forces on the lattice points. We have found that a non-zero $\th$ facilitates polaron propagation, in the sense that it lowers the critical amplitude $\Ac$, for any given $\bareps$. Moreover, a non-zero $\th$ results in a gradual onset of polaron motion, meaning the rate of change of polaron velocity with respect to $A$ near $A=\Ac$ is small, compared to the onset under $\th = 0$. Our results have also shown that, whenever there is polaron propagation, whether $\th = 0$ or $\th > 0$, the relative displacements between neighbouring lattice points remain under $\mcal{O}(10^{-2})$. This is a necessary condition which allows us to model the lattice points as point dipoles. 

Some of the choices of parameters in the MSPF may be justified physically as follows. It is well known that across the plasma membrane of a living cell, a \emph{resting membrane potential} is maintained by intercellular chemical processes \cite{Luckey2014}. It is also well known that within the plasma membrane there exist highly stable transmembrane regions of proteins, for instance the human prolactin receptor 2N7I \cite{Bugge2016}, and the rat monoamine oxidase A 1O5W \cite{Ma2004}, both of which are $\a$-helical structures spanning the entire membrane width. Given a constant potential difference of $\De V$ across a linear, homogeneous, isotropic dielectric medium with constant width $d$, the effective electric field inside the medium is given by $E_0 = \De V / (\k d)$, where $\k$ is the dielectric constant (a.k.a. relative permittivity) of the medium \cite{Jackson1999}. Assuming the plasma membrane is such a medium, we can then attribute the physical origin of $\bareps$ to the resting membrane potential, and calculate $\bareps$ using \cref{dimlessparams}, namely $\bareps = qE_0 R / (\hb \O)$. For many cells the values of the $E_0$ and $d$ are well established. As an example, one may look at human red blood cells (erythrocytes), one of the most widely studied cells in nature, and point to \cite{Cheng1980} for the value $E_0 = -8.4$mV, as well as \cite{McCaughan1980,Hochmuth1983} for $d = 78\tn{\ang}$. However, the value of $\k$ for a membrane is highly contentious, due to the fact that it depends sensitively upon a large variety of biophysical attributes of the membrane, such as hydration \cite{Mobley2008}, pH value \cite{Spassov2008}, and structural stability \cite{Vicatos2009}. In a recent review, it was reported that the value of $\k$ in literature ranges from 1 to 40 \cite{Li2013}. Feeding these values of $E_0, d$ and $\k$ into the equation for $\bareps$, we find that $\bareps$ ranges from 0.0033 to 0.13. We have taken care to ensure that in this study the values of $\bareps$ falls strictly within this range. For a physical origin of the periodic term, $A \sin (2\pi \t / T)$, one could look to common electromagnetic radiations which fill the environment around us in the modern age, such as the radiation from telecommunication transmitters. In particular, the values of $T$ which we have considered, 100, 500 and 2000, respectively match the frequencies of the IEEE 802.11ad protocal Wi-Fi band, the K\tsubs{u} band frequencies for satellite communications and broadcasting, and the UHF band frequencies for cellular communications \cite{IEEE80211ad,IEEE5212002}. However, the amplitudes of the aforementioned radiations are much smaller than the values of $A$ for which we have observed noteworthy results. For instance, treating the mobile telephone transmitter as an omni-directional dipole with peak power $P$, we can estimate the amplitude $\tilde{A}$ of its output waves at operational distance $d$, by using the well-known formula $P / (4 \pi d^2) = \eps_0 c \tilde{A}^2 / \k$, where $\eps_0, c, \k$ are the vacuum permittivity, speed of light, and relative permittivity of the medium, respectively. Feeding $P = 1\tn{W}$ \cite{Lonn2004} and $\k \le 40$ \cite{Li2013}  into the formula, we obtain dimensionless $A \le 6\times 10^{-6} / (d / \tn{metres})$. This means that in order to obtain $A = 0.1$, one needs the operational distance $d$ to be $\mcal{O}(10^{-5})$ metres, which is unrealistic. It is therefore clear that, in a real cell environment, the effect on a polaron due to a combination of resting membrane potential and random thermal forces are dominant over any external electromagnetic radiation that may commonly be present. This highlights the importance of our observation that, given a constant electric field $\bareps$, there exists some critical temperature $\thc$, such that the polaron undergoes directed drift if and only if $\th \ge \thc$. In other words, a combination of $\bareps$ and $\th$ can be sufficient for displacing the polaron, in a manner which causes minimal energy loss and which is directed, just as a combination of $\bareps$ and $A \sin(2 \pi \t / T)$ can. It has been reported that stationary polarons formed on 3-dimensional lattices, such as an \ahelix, are more strongly bound and therefore can be stabler during propagation \cite{Brizhik2004}. It is our hope that our model can be adapted to study such 3-D systems, and that the stabilising effect of the helical geometry will enable us to raise $\th$, from our current values of $\mcal{O}(10^1)$K to physiological temperatures. 

\section*{Acknowledgement}
We would both like to thank Larissa Brizhik for kindly answering some 
questions. 

\bibliographystyle{unsrt}

\end{document}